\documentclass[pdflatex,sn-nature]{sn-jnl}% Style for submissions to Nature Portfolio journals
\usepackage{graphicx}%
\usepackage{multirow}%
\usepackage{amsmath,amssymb,amsfonts}%
\usepackage{amsthm}%
\usepackage{mathrsfs}%
\usepackage[title]{appendix}%
\usepackage{xcolor}%
\usepackage{textcomp}%
\usepackage{manyfoot}%
\usepackage{booktabs}%
\usepackage{algorithm}%
\usepackage{algorithmicx}%
\usepackage{algpseudocode}%
\usepackage{listings}%

\usepackage{subfig}
\usepackage{subcaption}
\usepackage{lineno}
\theoremstyle{thmstyleone}%
\theoremstyle{thmstyletwo}%

\theoremstyle{thmstylethree}%

\begin{document}

\title[]{A unified cross-attention model for predicting antigen binding specificity to both HLA and TCR molecules}


\author[1]{\fnm{Chenpeng} \sur{Yu}}

\author[1]{\fnm{Xing} \sur{Fang}}
\author[1]{\fnm{Shiye} \sur{Tian}}
\author*[1]{\fnm{Hui} \sur{Liu}}\email{hliu@njtech.edu.cn}

\affil[1]{\orgdiv{College of Computer and Information Engineering}, \orgname{Nanjing Tech University}, \orgaddress{\city{Nanjing}, \postcode{211800 }, \state{Jiangsu}, \country{China}}}

%\affil[2]{\orgdiv{Department}, \orgname{Organization}, \orgaddress{\street{Street}, \city{City}, \postcode{10587}, \state{State}, \country{Country}}}

%\affil[3]{\orgdiv{Department}, \orgname{Organization}, \orgaddress{\street{Street}, \city{City}, \postcode{610101}, \state{State}, \country{Country}}}


\abstract{The immune checkpoint inhibitors have demonstrated promising clinical efficacy across various tumor types, yet the percentage of patients who benefit from them remains low. The bindings between tumor antigens and HLA-I/TCR molecules determine the antigen presentation and T-cell activation, thereby playing an important role in the immunotherapy response. In this paper, we propose UnifyImmun, a unified cross-attention transformer model designed to simultaneously predict the bindings of peptides to both receptors, providing more comprehensive evaluation of antigen immunogenicity. We devise a two-phase strategy using virtual adversarial training that enables these two tasks to reinforce each other mutually, by compelling the encoders to extract more expressive features. Our method demonstrates superior performance in predicting both pHLA and pTCR binding on multiple independent and external test sets. Notably, on a large-scale COVID-19 pTCR binding test set without any seen peptide in training set, our method outperforms the current state-of-the-art methods by more than 10\%. The predicted binding scores significantly correlate with the immunotherapy response and clinical outcomes on two clinical cohorts. Furthermore, the cross-attention scores and integrated gradients reveal the amino-acid sites critical for peptide binding to receptors. In essence, our approach marks a significant step toward comprehensive evaluation of antigen immunogenicity.}




\maketitle
\section{Introduction}\label{sec:introduction}
Immune checkpoint inhibitors have already demonstrated effective clinical antitumor efficacy in various types of tumors~\cite{schumacher2015neoantigens}. However, the percentage of patients who benefit from immunotherapy remains limited. A number of studies have confirmed that tumor antigens (neoantigens) play a crucial role in the immunotherapy response~\cite{glanville2017identifying,zhang2019combination}. In fact, the anticancer immune response involves a sequence of intricate biological events that lead to effective killing of cancer cells. Initially, tumor antigens are released by cancer cells through specific mechanism, and are captured and processed by antigen-presenting cells (APCs) \cite{yang2023antigen}. These APCs present the neoantigens on their outer cell surface (antigen presentation). Only if naive T cells recognize the antigenic epitopes and bind to the pHLA complex can they be conditionally activated and subsequently differentiate into effector T-cells, such as cytotoxic T lymphocytes (CTLs) \cite{barry2002cytotoxic,raskov2021cytotoxic,weigelin2021cytotoxic,rowen1996complete,glanville2017identifying}. The effector T-cells migrate to the tumor site and attack cancer cells \cite{fowell2021spatio}, ultimately inducing their death \cite{lim2020tumor}. These steps are referred to as the Cancer-Immunity Cycle~\cite{chen2013oncology}, which indeed highlights a delicate balance between the recognition of non-self antigens and the prevention of autoimmunity. Within this process, antigen presentation and T cell activation stand out as two steps critical to the success of the anticancer immune response \cite{yewdell1999immunodominance,dunn2004three}.

The binding of peptides to human leukocyte antigen (HLA) class I molecules is a fundamental step for neoantigen presentation \cite{yewdell1999immunodominance}. HLA alleles are well-known for their high specificity and polymorphism in the human population \cite{trowsdale2005hla}, leading to the restrictive binding of a narrow range of peptides~\cite{huppa2010tcr}. Following antigen presentation, the recognition of the presented antigens by T-cell receptors (TCR) is another crucial step to elicit T cells~\cite{huppa2010tcr}. This step is also highly selective, allowing only a small portion of antigens can be recognized and bound by TCRs. This selectivity, known as TCR binding specificity, arises from the high diversity of TCR repertoire (estimated to range from 10$^{15}$ to 10$^{61}$ possible receptors in humans)~\cite{nikolich2004many}. This diversity is primarily manifested in the complementarity determining region 3 (CDR3) \cite{zhang2016direct}, which directly interacts with the pHLA complex and determines the TCR binding specificity \cite{davis1988t,krogsgaard2005t}. The binding specificity ensures that only the immunogenic neoantigens can trigger immune response, thereby maintaining the delicate balance between effective immune responses and autoimmune reactions.

The HLA polymorphism and TCR diversity represent the evolutionarily acquired traits that enable the human immune system to respond to a wide array of pathogens at individual level \cite{chowell2019evolutionary,krishna2020genetic}. Some experimental assays like mass spectrometry (MS)-eluted HLA ligands~\cite{purcell2019mass}, and techniques such as single-cell TCR sequencing~\cite{zhang2018high} and T-scan~\cite{kula2019t} have been developed to detect pHLA and pTCR bindings, respectively. However, these experimental assays are often time-consuming, technically complex, and costly. To address these challenges, some computational methods have emerged as viable alternatives to predict peptide-receptor bindings~\cite{hudson2023can}. The pHLA prediction methods include TransPHLA~\cite{TransPHLA2022}, MHCflurry~\cite{2017MHCflurry}, NetMHCpan4.0~\cite{NetMHCpan}, DeepLigand~\cite{2019DeepLigand}, BERTMHC~\cite{2021BERTMHC}. The pTCR prediction methods include PanPep~\cite{PanPep2023}, pMTnet~\cite{pMTnet2021}, DLpTCR~\cite{2021DLpTCR}, ERGO2~\cite{ERGO22021}, TITAN~\cite{2021TITAN} and ATMTCR~\cite{fang2022attention}.  The ImmRep 2022 TCR-epitope specificity workshop released a dataset to benchmark the performance of more than ten predictive methods for pTCR bindings \cite{meysman2023benchmarking}. Although current methods have demonstrated promising predictive accuracy, they primarily focus on the prediction of pHLA or pTCR binding alone. However, the immunogenicity of antigens is actually influenced by the binding affinity to both HLA and TCR molecules, rather than just one.

\begin{figure}[hb]
        \centering
   \begin{minipage}{1\textwidth}   
        \centering
        \subfloat{
        \includegraphics[width=1\textwidth]{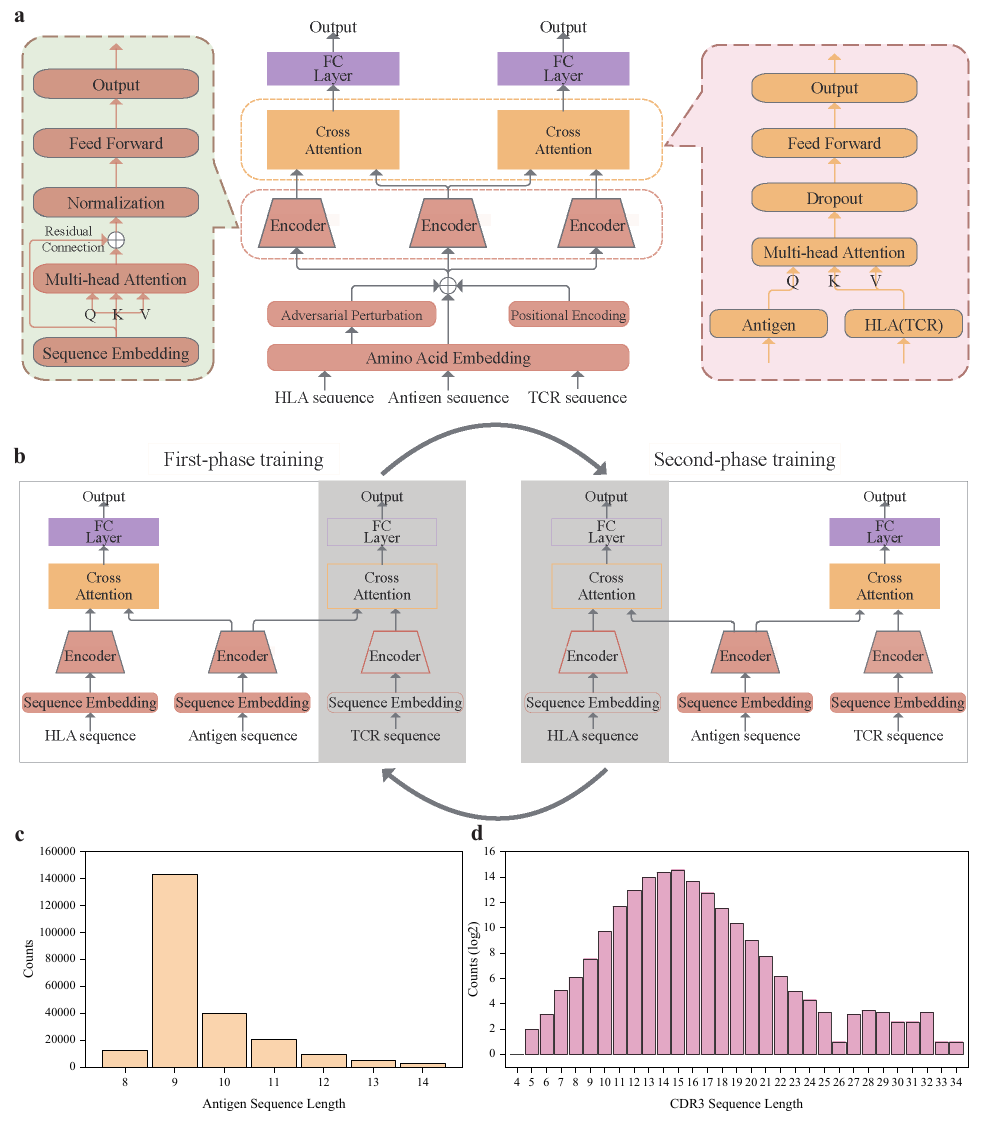}}
    \end{minipage}
       \caption{Illustrative diagram of UnifyImmun framework and two-phase training strategy, as well as the sequence frequency distributions of the benchmark datasets. (a) Architecture of UnifyImmun based on cross-attention mechanism. (b) Two-stage progressive training strategy. (c-d) Frequency of antigen sequences and TCR CDR3 sequences included in our created benchmark datasets with respect to lengths. }  \label{fig:framework}
\end{figure}

Distinct from previous studies that consider pHLA or pTCR binding specificity alone, we propose a unified model UnifyImmun, which integrates the predictive tasks of pHLA and pTCR bindings to establish a one-stop deep learning framework for comprehensive evaluation of antigen immunogenicity. UnifyImmun comprises three blocks: sequence embedding, encoder and cross-attention (Fig.\ref{fig:framework}a). The sequence embedding block receives the HLA, peptide, and TCR sequences as inputs, and respectively maps them into embeddings in latent space. Three self-attention encoders share common network structure, but operate with independent parameters to extract expressive features from three types of sequence embeddings, respectively. Next, two cross-attention layers are introduced to effectively fuse the features of the peptide-HLA pairs and peptide-TCR pairs, respectively. The output of the cross-attention layers is passed through fully-connected layers and softmax transformation to yield predictions for pHLA and pTCR binding, respectively. Particularly, the cross-attention scores offer valuable insights into the crucial amino acids and positional preference in peptides for binding to HLA and TCR molecules (see Result~\ref{sec:heatmap}). 

Given the vast diversity of HLA and TCR repertoires, the experimentally validated bindings currently available are limited and even biased, posing a tough challenge of overfitting in the development of prediction model. To overcome this limitation, we have introduced virtual adversarial training as a means to improve the model generalizability  (see Method~\ref{sec:adversarial}). Specifically, we apply adversarial perturbations to the sequence embeddings to generate virtual adversaries that aim to maximize the loss function. The adversarial training makes our model less sensitive to slight changes in input sequences, thereby significantly improves the performance (see Method~\ref{sec:ablation}).

Ideally, our model prefers to be trained using HLA-peptide-TCR triplet samples. However, the availability of such triplets is currently limited, whereas pHLA and pTCR pairwise bindings are relatively abundant (Fig.~\ref{fig:framework}c,d; Supplementary Figure 1). To efficiently leverage the available data, we propose a two-stage progressive training strategy (Fig.~\ref{fig:framework}b). Through performance evaluation on multiple test sets, we have demonstrated that the two-stage training effectively enhances the feature extraction capabilities of the encoders, thereby improving the performance in predicting pHLA and pTCR binding specificity.

Our model exhibits advantages over previous methods on both pHLA and pTCR binding prediction tasks, and offers at least three notable contributions as follows:
\begin{itemize}
    \item By integrating both prediction tasks into a unified model, our method enables simultaneous evaluation of the potential in predicting pHLA and pTCR bindings. Such two-faceted assessments provide a more holistic view of antigen immunogenicity than previous methods, offering a new insight into the neoantigen quality for triggering immune response. Moreover, once trained, our model can be independently applied to three prediction tasks: pHLA binding, pTCR binding, and HLA-peptide-TCR binding. 
    \item We devise a two-phase progressive training strategy to make full use of the pHLA and pTCR pairwise binding data available. Our experiments have validated that these two tasks mutually improve each other, by compelling the encoders to extract more expressive features. Furthermore, the virtual adversarial training effectively enhances the model generalizability. 
    \item Our extensive experiments on multiple independent and external test datasets have verified that the unified model achieved superior performance over current state-of-the-art methods on both prediction tasks. Moreover, the cross-attention scores facilitated the capture of underlying patterns of peptides binding to HLA and TCR molecules. 
\end{itemize}

\section{Results}\label{sec:results}
\subsection{Performance evaluation on pHLA binding prediction}
\begin{figure}[hb]
         \begin{minipage}{1\textwidth}
        \centering
        \subfloat{\includegraphics[width=1\textwidth]{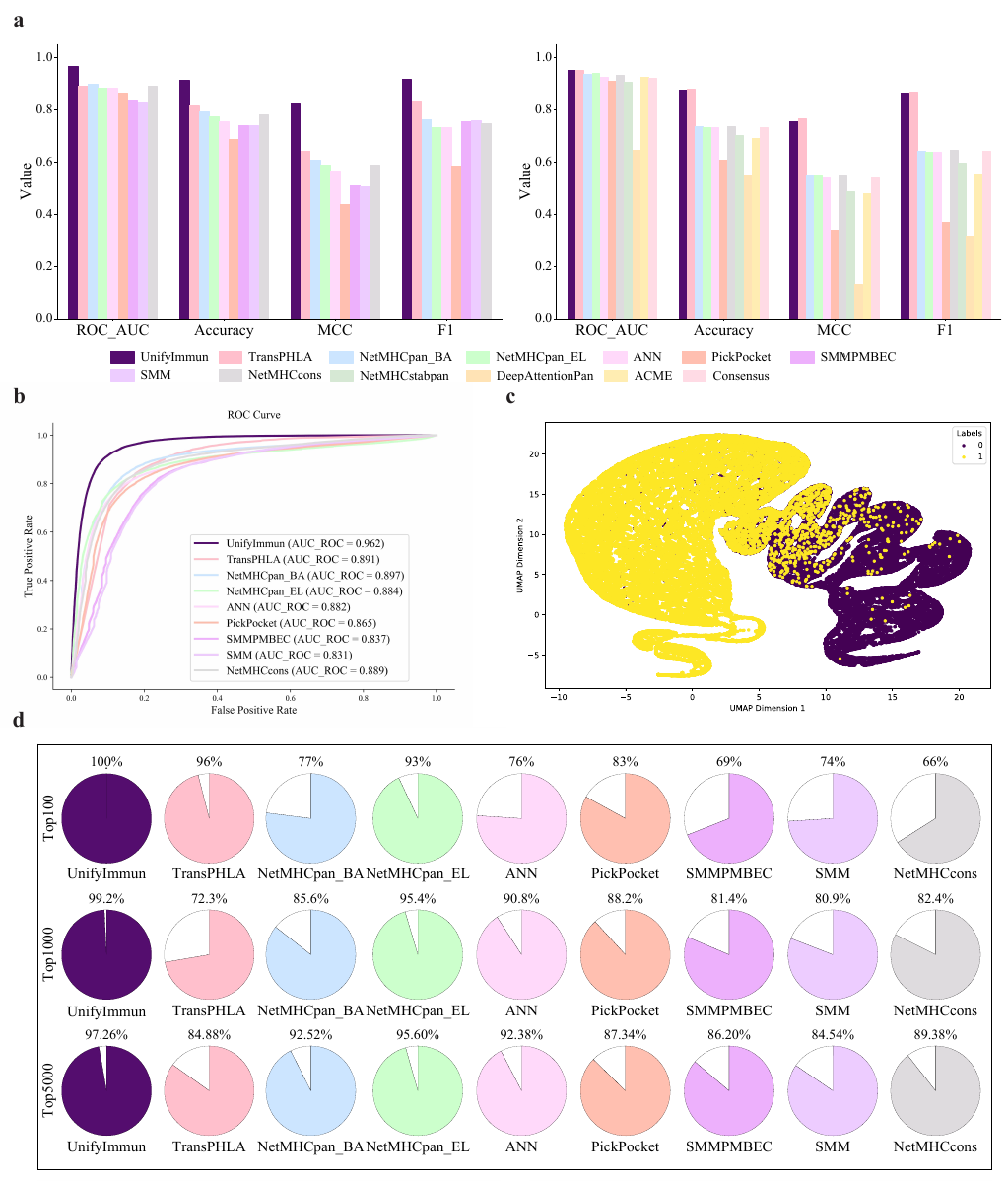}}
    \end{minipage}
    \caption{Performance evaluation on predicting peptide-HLA binding specificity. (a) Performance comparison to twelve existing methods on independent (left) and external (right) test dataset, respectively. (b) ROC curves and AUC values achieved by UnifyImmun and eight competing methods on hold-out independent test set. (c) UMAP feature visualization of peptide-HLA pairs. (d) Positive predictive value (PPV) for the top 100, top 1000, and top 5000 predicted pHLA samples. }  \label{fig:hla-antigen}
\end{figure}

To evaluate our model performance in predicting pHLA binding, we conducted performance assessment on four datasets: independent test, external test, HPV and neoantigen validation datasets. We compared UnifyImmun against twelve established methods, including NetMHCpan\_EL \cite{NetMHCpan}, NetMHCpan\_BA \cite{NetMHCpan}, ANN \cite{ANN}, PickPocket \cite{PickPocket}, SMMPMBEC \cite{SMMPMBEC}, SMM \cite{SMM},NetMHCcons \cite{2012NetMHCcons}, NetMHCstabpan \cite{Rasmussen2016Pan} and Consensus \cite{2006A}, as well as three recently published attention-based methods, TransPHLA \cite{TransPHLA2022}, ACME \cite{ACME} and DeepAttentionPan \cite{2021Deep}. These competing methods were downloaded as executable packages and run on the same test sets using their recommended parameters. We reported multiple performance metrics, such as AUROC, accuracy, MCC, and F1-score (Fig.~\ref{fig:hla-antigen}a). To offer more comprehensive evaluation, we also provided other metrics , including precision, recall, AUPR, and specificity (Extended Data Figure 1a).

We first evaluated the performance of UnifyImmun against other competing methods on an independent set, which contained 10\% pHLA samples held out from our established benchmark dataset (see section \ref{sec:dataset}). The results showed that UnifyImmun remarkably outperformed all other methods across all evaluation metrics (Fig.~\ref{fig:hla-antigen}a). In particular, compared to the second-best method, TransPHLA, UnifyImmun achieved at least a 5\% improvement in both AUROC and AUPR. To provide a visual presentation of the performance differences, we presented the ROC curves (Fig.\ref{fig:hla-antigen}b) and precision-recall curves (Extended Data Fig.1b) for all competing methods on the hold-out test set. These curves further validated the superior performance of our proposed method. The UMAP feature visualized that the positive and negative samples separated notably in the latent space (Fig.~\ref{fig:hla-antigen}c). To further assess the model's ability to prioritize pHLA bindings, we presented the positive predictive value (PPV) for the top 100, top 1000, and top 5000 predicted positive samples (Fig.~\ref{fig:hla-antigen}d). UnifyImmun achieved an impressive 100\% PPV for the top 100 predictions and maintained excellent performance above 97\% for both the top 1000 and top 5000 predictions. In contrast, the other methods did not demonstrate comparable ability in prioritizing pHLA bindings. We also paid attention to the top five HLA alleles with most bindings, including HLA-B27:05, HLA-A02:01, HLA-A03:01, HLA-B07:02, and HLA-B15:01. Our model achieved the best performance on the HLA-A03:01 allele bound by 8-mer peptides ($n$=385) with 0.941 AUROC value, followed by 9-mer peptides ($n$=17,658) with 0.915 AUROC value (Extended Data Table 1). The performance was relatively poor on the HLA-B27:05 allele bound by 14-mer peptides. We also separately checked the performance on the different test subsets split by peptide lengths, and found that UnifyImmun gained the highest AUROC values 0.91 for 9-mer and 11-mer peptides, and the lowest AUROC value 0.88 for 14-mer peptides (Extended Data Table 2).

For objective evaluation, we conducted performance comparison experiments on an external pHLA binding dataset provided by Anthem~\cite{Anthem}. This external set included 103,854 pHLA bindings that cover 5 HLA alleles and 100,581 distinct peptides, with approximately balanced numbers of positive and negative samples. While TransPHLA exhibited performance advantages on this dataset, UnifyImmun achieved nearly identical performance compared to TransPHLA (Fig.~\ref{fig:hla-antigen}a) across all performance metrics. Furthermore, UnifyImmun notably outperformed all other methods except for TransPHLA. 

The HPV dataset came from a previous study \cite{bonsack2019performance} that identified 278 experimentally verified pHLA bindings derived from the HPV16 proteins E6 and E7, consisting of peptides ranging from 8 to 11 amino acids in length \cite{wells2020key,wang2020ineo}. We compared UnifyImmun against fifteen previous methods on this test set. Because some competing methods cannot accommodate every HLA allele and peptide length, thereby failed to cover all test samples. UnifyImmun achieved an impressive accuracy rate of 83.8\% (Extended Data Figure 1c). This significantly surpassed the performance of the second-best model TransPHLA, which only correctly identified 68\% pHLA bindings. The neoantigen validation dataset includes 221 experimentally verified pHLA bindings~\cite{TransPHLA2022}, which were collected from non-small-cell lung cancer, melanoma, ovarian cancer and pancreatic cancer in recent studies. On this test set, UnifyImmun achieved 94.1\% accuracy (correctly identifying 208 out of 221) and actually performed comparable to TransPHLA with 96.4\% (Extended Data Figure 1d). These different types of test sets were complementary in the performance evaluation, so only a method that worked well on all the test sets can demonstrate its superiority. Collectively, the performance comparison experiments on four distinct datasets clearly demonstrated the superior generalization capability of UnifyImmun over previous methods.

\subsection{UnifyImmun boosts predictive performance of pTCR bindings}
To assess the performance in predicting pTCR binding specificity, we conducted performance comparison with four current state-of-the-art methods devised for the task, including PanPep~\cite{PanPep2023}, ERGO2~\cite{ERGO22021}, pMTnet~\cite{pMTnet2021} and DLpTCR~\cite{2021DLpTCR}. For PanPep, ERGO2, and pMTnet, we executed their executable codes using their recommended parameters on the same workstation as UnifyImmun. For DLpTCR, we accessed its web server to obtain its predicted results of the test sets.  We observed significant differences in the inference efficiency among the evaluated methods. Upon comparing time overhead on our local workstation for four methods (excluding DLpTCR), UnifyImmun exhibited markedly higher inference efficiency compared to PanPep and pMTnet, while being slightly less efficient than ERGO2 (Supplementary Figure 6). 

\begin{figure*}[hb]
        \begin{minipage}{1\textwidth}
        \centering
        \subfloat{\includegraphics[width=1\textwidth]{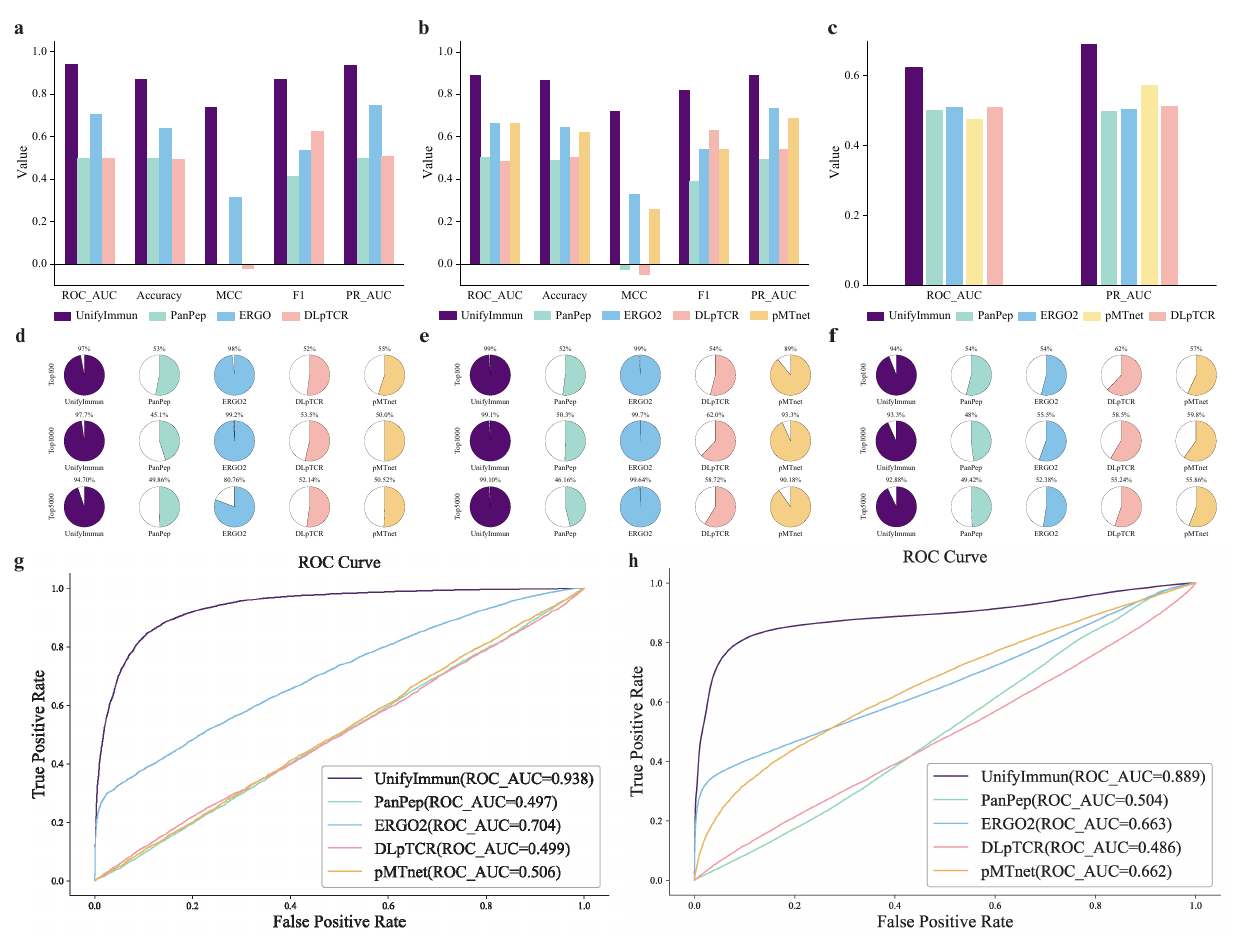}}
    \end{minipage}
    \caption{Performance evaluation on predicting peptide-TCR binding specificity. (a-c) Performance comparison to four methods on independent, external, and COVID-19 test sets, respectively. (d-f) Positive predictive value (PPV) for the top 100, top 1000, and top 5000 predicted samples on independent, external, and COVID-19 test sets, respectively. (g-h) ROC curves and AUC values on independent and external test dataset, respectively.  }  \label{fig:tcr-antigen}
\end{figure*}

We initially evaluated the performance of these methods on our established pTCR binding dataset with negative samples generated through random mismatching. On the 10\% hold-out independent test set (see section \ref{sec:dataset}), we found that UnifyImmun remarkably outperformed all other methods (Fig.~\ref{fig:tcr-antigen}a,g; Extended Data Figure 2a,c). Specifically, UnifyImmun achieved AUROC and AUPR values of 0.938 and 0.936, highlighting its exceptional predictive ability in predicting pTCR binding specificity. Among the competing methods, only ERGO2 exhibited moderate performance, with AUROC and AUPR values of 0.704 and 0.747, respectively. Other methods performed close to random guessing, indicating their weak generalizability in predicting pTCR bindings. For specific length peptide, UnifyImmun achieved the highest AUC value of 0.95 for 9-mer peptides, and the lowest AUC value of 0.87 for 12-mer peptides (Extended Data Table 3).

For further evaluation, we compiled an external test set of pTCR binding pairs from a number of publications, which included 97,043 pTCR pairs spanning 1,239 distinct peptides and 24,856 CDR3 sequences. This external set did not contain any shared peptide with the training set, allowing us to assess the predictive capacity of UnifyImmun toward real-world scenarios beyond the hold-out test set. As expected, UnifyImmun achieved AUROC values of 0.889 and AUPR values of 0.888, significantly outperforming the second-best method, ERGO2, which obtained only about 0.663 AUROC and AUPR values (Fig.~\ref{fig:tcr-antigen}b,h; Extended Data Figure 2b,d). The other three methods exhibited even poorer performance on the external set. We observed even negative MCC values for PanPep and DLpTCR, indicating a high degree of disagreement between their prediction and ground truth. This observation reflected the limitations of the previously published methods, and in turn validated that UnifyImmun achieved superior performance in predicting pTCR binding specificity between unseen peptides and TCR sequences.

To check the ability to prioritize pTCR bindings, we computed the positive predictive value (PPV) for the top-ranked predicted pTCR samples on the two distinct datasets mentioned above. Specifically, we evaluated the PPV for the top 100, top 1000, and top 5000 predictions (Fig.~\ref{fig:tcr-antigen}d,e). UnifyImmun achieved an impressive 97\%, 97.7\% and 94.7\% PPV values for the top 100, top 1000, and top 5000 predictions, respectively. In contrast, the prioritization ability of the other methods was inferior to UnifyImmun. The UMAP feature visualization of pTCR pairs implied that the positive and negative samples separated remarkably in the latent space (Supplementary Figure 3a-b). It is indeed noteworthy that while the competing methods may exhibit promising results on small datasets, their performance decreased seriously when tested on large-scale dataset. This implies they suffer from weak generalization ability and struggle to adapt to large real-world data scenarios. In contrast, UnifyImmun demonstrated strong robustness across distinct datasets, offering a more dependable and precise tool for predicting pTCR binding specificity. 

Moreover, we employed an alternative strategy to generate negative pTCR samples (see Methods for details). This strategy combined random mismatching and an unbound sequence pool, with each contributing 50\% of the negative pTCR samples. Following this, we conducted performance evaluation experiments similar to those described above. The experimental results included the performance metrics of our method alongside four competing methods on both the hold-out independent test set and the external test set (Supplementary Figure 4). While all methods exhibited a decline in performance compared to the dataset with negative samples generated exclusively through random mismatching, our method consistently outperformed the four competing algorithms across various performance metrics, including the PPV for the top 100, 1000, and 5000 predicted pTCR bindings. The results strongly demonstrated that our method achieved significantly better generalizability across different datasets whose negative samples were generated using different strategies.

\subsection{Two-phase progressive training improve model performance}
\begin{figure*}[hb]
        \begin{minipage}{1\textwidth}
        \centering
        \subfloat{\includegraphics[width=1\textwidth]{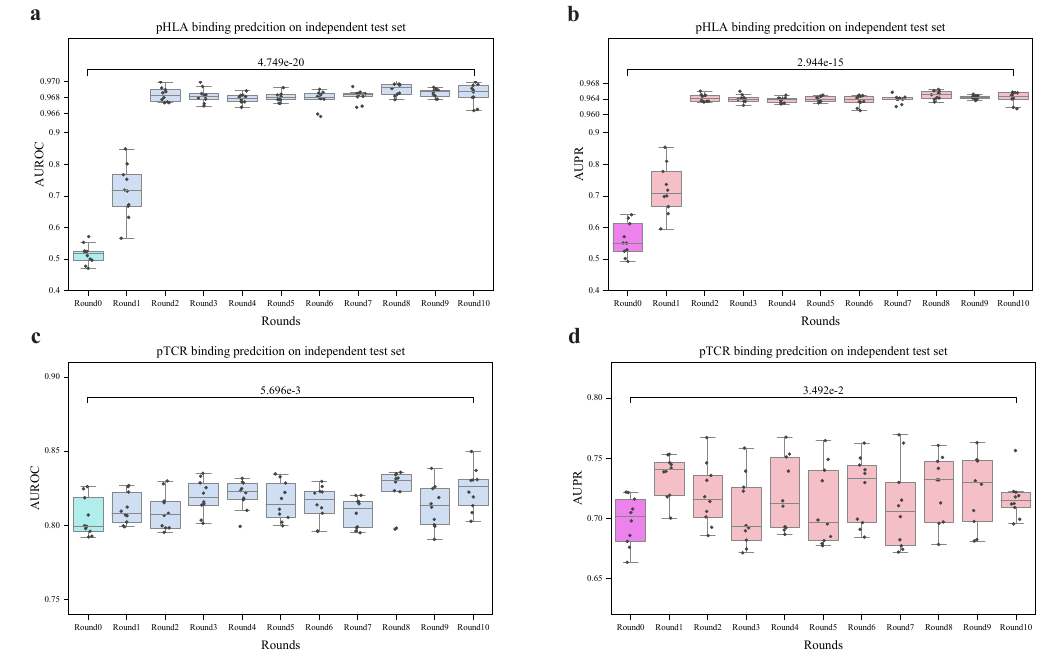}}
    \end{minipage}
    \caption{Two-phase progressive training improved performance for both pHLA and pTCR binding prediction tasks. (a-b) AUROC and AUPR values increased with two-phase training rounds on pHLA independent test set. (c-d) AUROC and AUPR values increased with two-phase training rounds on the pTCR independent test set.}  \label{fig:two-stage-training}
\end{figure*}

Due to the limited number of HLA-antigen-TCR triplet samples for model training, we devised a two-phase progressive training strategy (see section~\ref{sec:twophase}) aimed at effectively leveraging the available pHLA and pTCR pairwise binding data. 
To validate the performance enhancement from two-phase training, we randomly divided our established benchmark datasets into training and test sets for model training and subsequent performance assessment. This process was independently repeated ten times to account for variations introduced by random data partitioning. We presented the results in boxplots for each training round (Fig.~\ref{fig:two-stage-training}). 
The results showed that the performance of our method was suboptimal for both pHLA and pTCR binding prediction in the absence of alternating training, namely, Round 0. As the number of training rounds progressed, we noticed a significant and steady increase in performance until it stabilized at a notably high level. Specifically, on the pHLA hold-out independent test set, the two-phase training quickly boosted the AUROC and AUPR values (Fig.~\ref{fig:two-stage-training}a,b). The one-way analysis of variance (ANOVA) revealed statistically significant differences between the first and last bins for the AUROC and AUPR values ($F$-test, $p$-value= 4.75e-20 and 2.94e-15, respectively).

For pTCR binding prediction, similar trends can be observed on the independent test set (Fig.~\ref{fig:two-stage-training}c,d). As expected, the increasing number of training rounds improved the model's performance on the independent test set. The ANOVA analysis confirmed the statistically significant differences between the first and last bins for the AUROC and AUPR values ($F$-test, p-value= 0.0167 and 0.0395, respectively).  Notably, we observed that the variance of the performance metrics was relatively high in the early rounds, it decreased progressively as the number of training rounds increased. This indicated that the model performance became less affected by random data partitioning as training progressed. These results confirmed that the two-stage progressive training strategy effectively drove the encoders to learn more informative features, thereby achieving more reliable performance.

\subsection{Cross-attention scores reveal critical peptide sites} \label{sec:heatmap}
\begin{figure*}[hb]
        \begin{minipage}{1\textwidth}
        \centering
        \subfloat{\centering\includegraphics[width=0.95\textwidth]{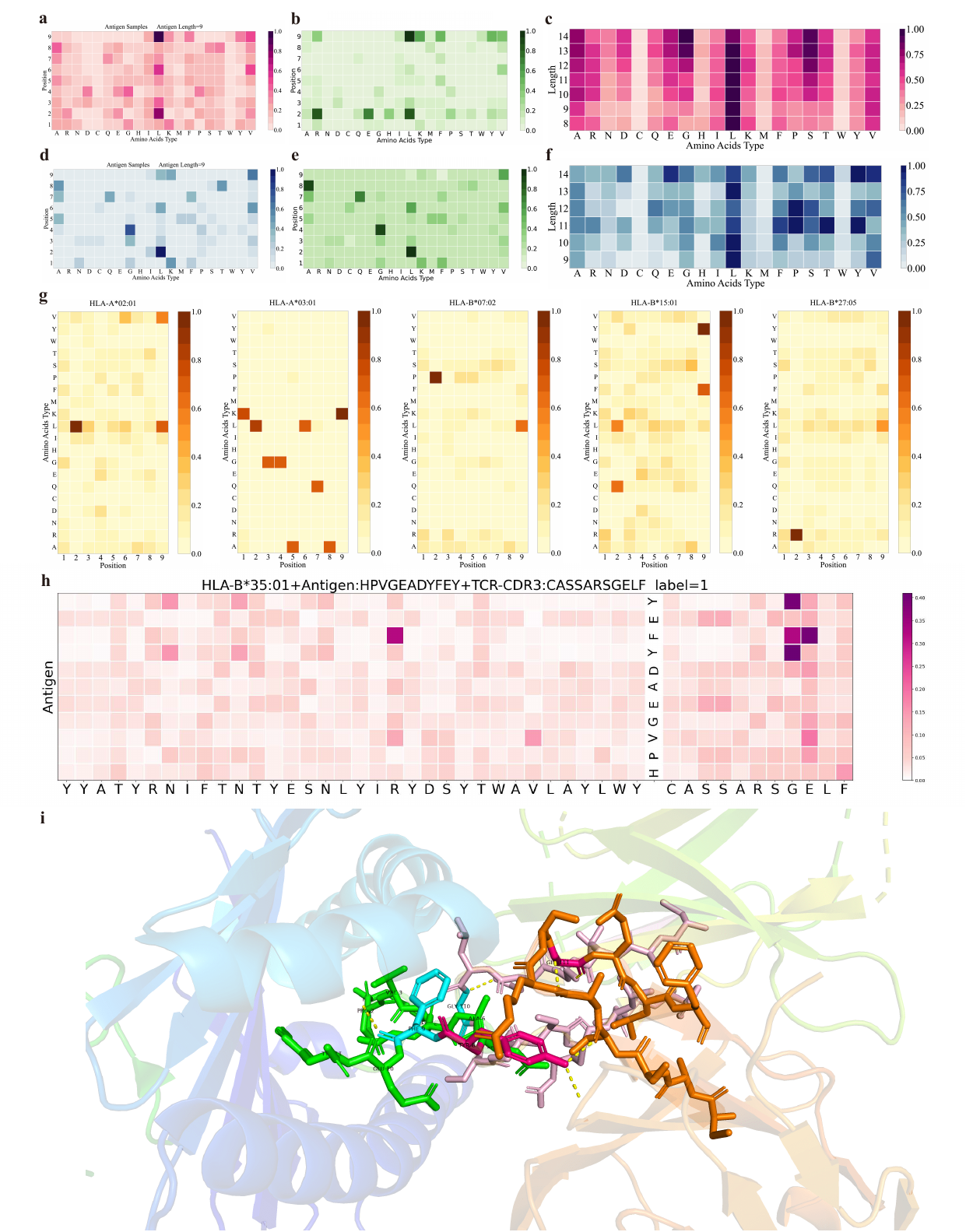}}
    \end{minipage}
    \caption{Heatmaps generated from cross-attention scores and integrated gradients. (a-b) Heatmaps of cross-attention scores and integrated gradients of the amino-acid type at each position of 9-mer peptide binding to HLA molecules. (c,f) Accumulative attention scores across peptide length of each amino-acid type of peptide binding to HLA and TCR molecules, respectively. (d-e) Heatmaps of cross-attention scores and integrated gradients of the amino-acid type at each position of 9-mer peptide binding to TCR molecules. (g) Heatmaps of cross-attention scores for top five HLA alleles with most 9-mer binding peptides. (h-i) Attention score-based heatmap and 3D structure for TCR complex with HLA-B35:01/HPVG (PDB ID: 3MV7).
 }  \label{fig:locus}
\end{figure*}

We employed the cross-attention mechanism to integrate the features of peptides and HLA/TCR molecules, allowing us to explore whether cross-attention scores reflect the key positions and amino-acid types within the peptide that determine its binding affinity to HLA or TCR molecules. For this purpose, we aggregated the cross-attention scores for amino-acid type at every position across all peptide sequences. As a result, a higher score indicates its strong influence on the binding affinity to corresponding receptor. In order To accommodate peptides of different lengths, we independently generated the heatmaps for peptides ranging from 9 to 14 amino acids in length (Fig.~\ref{fig:locus} and Supplementary Figures 7-8).

Since 9-mer peptides are the most common, we inspected the heatmaps to uncover important amino-acid types and positions. For pHLA binding, we found remarkably higher attention scores at the second position (P2) and C terminus  (Fig.~\ref{fig:locus}a), indicating that these two sites make significant contributions to the peptides bound by HLA molecules. Meanwhile, the Leu (L) amino acid consistently received higher attention scores, especially at the two key positions, emphasizing its significance for pHLA binding. For pTCR binding, we also found that the Leu at the second peptide position \cite{parker1992sequence} stands out as the most influential (Fig.~\ref{fig:locus}d). 
To support the observations of cross-attention heatmaps, we calculated the Integrated Gradients (IG) for each amino-acid type at every position within the peptide (Supplementary Figures 9-10). For 9-mer peptides, the IG heatmaps exhibited high similarity to the attention heatmaps (Fig.~\ref{fig:locus}b,e). For instance, the Leu amino acid at the second position within the peptides emerged as crucial residue for the binding to both HLA and TCR molecules. We also observed high frequency of the amino acids in these peptide positions (Supplementary Figure 1d-e). For other length peptides, we also observed that the Leu amino acid at the P2 and C terminus is highlighted in the heatmaps generated from both attention and IG values. The consistency between the attention scores and the IG values reinforced the validity of our model and highlighted the crucial role of specific amino-acid types and their positions in mediating peptide binding affinity. Furthermore, we calculated the cumulative cross-attention scores for each amino-acid type across all positions within specific-length peptides. The cumulative value reflects the overall importance of specific amino-acid type in mediating peptide binding to HLA or TCR. We illustrated the heatmaps generated from the cumulative scores of 20 distinct amino acids in peptides ranging from 8 to 14 amino acids in length (Fig.~\ref{fig:locus}c,f). Clearly, Leu consistently demonstrated high importance in peptides binding to both types of receptors.

To explore the crucial residues in peptide binding to specific HLA alleles, we generated the attention heatmap for the top 5 HLA alleles with most 9-mer binding peptides, including HLA-A02:01, HLA-A03:01, HLA-B07:01, HLA-B15:01 and HLA-B27:05 (Fig.~\ref{fig:locus}g). It can be observed that the Leu amino acid at the second position significantly affects antigen binding to HLA-A02:01 and HLA-A03:01. For HLA-B15:01 and HLA-B27:05, the Tyr (Y) at the C terminus and Arg (R) at P2 exhibited avdominant role. In fact, several studies have reported the crucial amino acids and their positional preferences in peptide binding to specific HLA alleles. Our heatmaps confirmed these reported residues that exhibited significantly high scores at their preferential positions. For instance, Dibrino et al. \cite{dibrino1993hla,dibrino1994endogenous} showed that HLA-A1 prefers Asp(D)/Glu(E) at P3 and Tyr(Y) at the C terminus, which was consistently reflected in our heatmap for HLA-A01:01 (Extended Data Figure 3b). Similarly, other studies have described preferential amino acids for HLA-B8, HLA-B14, HLA-B27, and HLA-B44, all of which were well-represented in our attention heatmaps \cite{dibrino1994endogenous,dibrino1994hla,parker1994pocket,dibrino1995identification}. These findings strongly validate the crucial amino acids and their positional preferences in peptide binding to specific HLA alleles as reported by previous studies and uncovered by our methods.

For intuitive visualization of the important sites and amino acids involved in HLA-peptide-TCR binding, we obtained the crystal structure of the TK3 TCR in complex with HLA-B*3501/HPVG (PDB ID: 3MV7) from the PDB database. We extracted the HLA allele, antigen (11-mer), and the CDR3 $\beta$ chain (11-mer) and predicted the pairwise binding probabilities using UnifyImmun. The results indicated a high binding score between the HLA-B*3501 allele and antigen (0.99), as well as a moderate score between the CDR3$\beta$ chain and antigen (0.62). Furthermore, the attention heatmap for this pHLA-TCR complex (Fig.~\ref{fig:locus}h) revealed a significant cross-attention score between the 8-th amino acid Tyr (Y) of the antigen and the 8-th amino acid Gly (G) of the CDR3 $\beta$ chain. For this CDR3 $\alpha$ chain, we obtained similar attention heatmap (Extended Data Figure 3a). Upon careful inspection of the three-dimensional structure (Fig.~\ref{fig:locus}i), we found the hydrogen bonds associated with these two amino acids, indicating their crucial role in the formation of the pTCR complex, despite their distance of 8.77$\mathring{A}$ (distance between two C$\alpha$ atoms of two amino acids) is slightly beyond the conventional contact threshold of 6$\mathring{A}$. We also observed that the 9-th amino acid Phe (F) in the antigen received remarkably high cross-attention scores for HLA binding. Consistently, the complex crystal structure showed that Phe is embedded in the HLA binding groove (represented by blue helices) and stabilized through hydrogen bonds (yellow lines).  In addition, the heatmaps of two randomly selected pHLA-TCR complexes (one positive and one negative) revealed that the attention scores in the positive sample were significantly higher than those in the negative sample (Supplementary Figure 11a-b). In summary, the cross-attention mechanism offers an opportunity to explore the global dependencies between TCR-pHLA interactions, thereby enhancing the interpretability of our model.

%For an intuitive visualization of the key positions and amino acids involved in TCR-pHLA binding, we randomly selected 100 HLA-antigen-TCR samples (50 positive and 50 negative) and generated cross-attention heatmaps for pairwise combinations (Figure~\ref{fig:locus}(g-h) and Figure S4). Figures 6(g-h) illustrated the examples for one positive and one negative sample, respectively. Overall, we observed that the cross-attention scores were significantly higher in the positive samples compared to the negative one. As expected, the Leu amino acid at the second position of the antigen sequence exhibited a high attention score associated to the Arg amino acid in the HLA sequence. Also, the final Phe amino acid in the CDR3 sequence demonstrated high attention scores with multiple amino acid in antigens. These findings suggested that cross-attention scores provided our model with a certain degree of interpretability, allowing for a better understanding of the underlying mechanisms of TCR-pHLA binding.

\subsection{High generalizability to COVID-19 pTCR binding prediction} \label{sec:covid}
To validate the generalizability of UnifyImmun, we tested its ability to predict the bindings between COVID-19 virus-derived antigens and TCRs. We collected more than 540,000 bindings between antigens derived from COVID-19 virus and human TCRs from the ImmuneCODE$^{TM}$ database~\cite{nolan2020large}.
To demonstrate UnifyImmun's predictive capability for novel peptides, we removed all pTCR samples associated with the shared peptides in the training set, and ensured that all peptides in this test set were unseen in the model training stage. Meanwhile, we generated an equal number of negative samples via random mismatching, thereby creating a million-scale COVID-19 test set. It is worth noting that this is the largest pTCR binding test set to date. We compared UnifyImmun with several other methods, including PanPep, ERGO2, DLpTCR, and pMTnet. Due to the low efficiency of pMTnet and DLpTCR, they were unable to tackle the million-scale test set within a reasonable time frame. Therefore, we randomly selected 100,000 pairs as their test set for evaluation.

The results illustrated that UnifyImmun achieved an AUROC value of 0.623 (Fig.~\ref{fig:tcr-antigen}c, Supplementary Figure 5),. In contrast, other methods obtained AUROC values only slightly above 0.5, which is close to random guessing. Clearly, UnifyImmun outperformed all competitive methods by more than 10\% in this unseen peptide context. Furthermore, we computed the PPV values for the top 100, top 1000, and top 5000 predictions made by each method. Our model consistently achieved 90\% PPV values, remarkably outperformed all the competing methods whose PPV values were always less than 65\% (Fig~\ref{fig:tcr-antigen}f). 
In addition, we conducted a performance evaluation on the dataset containing negative pTCR samples generated using the hybrid strategy (Supplementary Figure 4c,f). The experimental result validated that UnifyImmun consistently outperformed four competing methods across various performance metrics, as well as the PPV for the top 100, 1000, 5000 predicted pTCR bindings. 
Overall, the significant advantages over previous methods strongly validated the robust generalizability of UnifyImmun, and highlighted its potential for facilitating the development of effective immune-based therapies and vaccines against COVID-19 viruses.

% \begin{figure*}[hb]
%     \begin{minipage}{1\textwidth}
%         \centering
%         \subfloat[AUROC and AUPR on COVID-19 test set ]{\includegraphics[width=0.43\textwidth]{figure/Fig6/TCR-Covid-Bar.pdf}}
%         \subfloat[PPV on COVID-19 test set]{\centering\includegraphics[width=0.57\textwidth]{figure/Fig6/PPV-TCR-Covid.pdf}}
%     \end{minipage}
%    \caption{Performance comparison to four existing methods on COVID-19 antigen-TCR binding test set. (a) AUROC and AUPR values on COVID-19 test set. (b)PPV of top100, top1000 and top5000 predicted samples on COVID-19 test set. }  \label{fig:Covid}
% \end{figure*}

\subsection{Predicted binding scores correlated immunotherapy outcomes}
The antigen presentation to cytotoxic T-cells plays a pivotal role in determining the efficacy of tumor immunotherapy, particularly with immune checkpoint inhibitors. To evaluate the predictive power of UnifyImmun, we conducted correlation analysis on two cancer cohorts: a metastatic melanoma cohort (MM-HLA) \cite{van2015genomic} for pHLA binding, and an advanced melanoma cohort~\cite{RIAZ2017934} for pTCR binding (MM-TCR). 

\begin{figure*}[hb]
    \begin{minipage}{1\textwidth}
        \centering
        \subfloat{\includegraphics[width=1\textwidth]{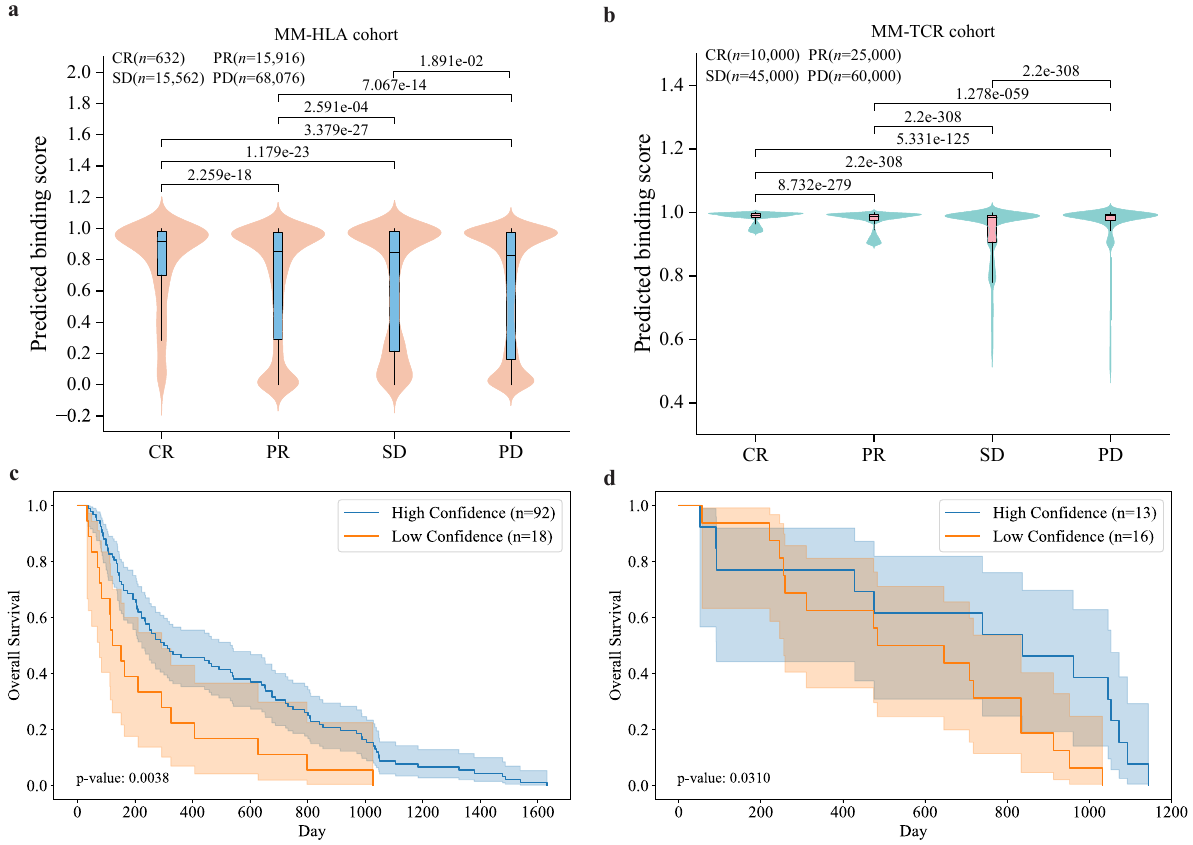}}
    \end{minipage}
    \caption{Correlation between UnifyImmun predicted binding scores and immunotherapy response and clinical outcomes on two clinical cohorts. (a-b) Violin plots of predicted pHLA and pTCR binding scores regarding the different immunotherapy response groups of MM-HLA cohort and MM-TCR cohort, respectively. (c-d) Survival curves between stratified patient groups with high- and low-confidence antigen binding specificity on MM-HLA and MM-TCR cohorts, respectively.}   \label{fig:clinical}
\end{figure*}

The MM-HLA cohort included 110 patients, with each patient harboring an average of 919 neoantigens. For each patient, we obtained the HLA typing, antigen sequences, immunotherapy responses, and clinical outcomes. For the MM-HLA cohort, we applied the RECIST criteria to categorize the patients into four groups: complete response (CR, $n$=3), partial response (PR, $n$=14), stable disease (SD, $n$=12), and progressive disease (PD, $n$=76). We predicted the binding probabilities for all possible HLA-peptide pairs using UnifyImmun and visually represented the predicted results for each patient group. The one-way analysis of variance (ANOVA) with an $F$-test revealed statistically significant differences in the pHLA binding affinity between these groups (Fig.~\ref{fig:clinical}a). Notably, the PD group demonstrated a highly statistically significant divergence compared to the other patient groups. This observation implied the differences in neoantigen presentation between benefit vs non-benefit patient groups from immunotherapy. Moreover, the patients in the CR group exhibited the prevalence of high-scored neoantigens, while those in the PD group showed many low-scored neoantigens. If the patients were stratified into response ($n$=27), non-response ($n$=73), and long survival ($n$=10) groups according to the standard that PFS is less than 180 days but OS is more than 2 years, they exhibited distinct patterns in antigen binding to HLA molecules (Fig.~\ref{fig:clinical}c; For details of the violin plots see Supplementary Figure 12e). As a contrast, we tested TranspHLA and netMHCcons on MM-HLA. The results showed that TranspHLA was not effective enough to statistically distinguish the pHLA bindings between CR and PR patients in the MM-HLA cohort, as well as between the CR and SD group (Extended Data Figure 4a, p-value=0.271 and 0.682, respectively).

The MM-TCR cohort comprised 29 patients who had received immunotherapy, and each patient underwent TCR-seq and genomic sequencing. Taking the amino acid resulting from a missense mutation as an anchor, we generated all possible 9-mer peptides harboring this anchor site. After extracting the CDR3 sequences from TCR-seq data, we created all possible peptide-CDR3 pairs for each patient, yielding a total of 81,851,486 pairs. We used UnifyImmun to score the pairs and selected the top 5000 highest-scoring pairs for each patient. Next, we categorized the patients into CR ($n$=2), PR ($n$=5), SD ($n$=9), and PD ($n$=12) groups, and plotted the boxplots of the predicted scores for each group (Fig.~\ref{fig:clinical}b). The one-way ANOVA analysis revealed statistically significant differences among the groups ($F$-test, Fig.~\ref{fig:clinical}b), with the CR and PR groups harboring pTCR pairs with significantly higher scores than the SD and PD groups. By stratifying the patients into benefit ($n$=13), non-benefit ($n$=13), and long-term survival groups ($n$=3), we found that the patients in the long-term survival group exhibited highly scored pTCR pairs compared to other groups (Supplementary Figure 12f). 

Finally, to confirm the correlation between highly scored pHLA and pTCR pairs by UnifyImmun and improved clinical outcomes, we conducted survival analysis on two melanoma cohorts (MM-HLA and MM-TCR). We considered the top 2\% pHLA and pTCR pairs as high-confidence bindings, and stratified the patients with such bindings into the high-confidence group, while the remaining patients were placed into the low-confidence group. The survival analysis showed that the high-confidence group exhibited significantly higher overall survival (OS) and progression free survival (PFS) compared to the low-confidence group (Fig.~\ref{fig:clinical}c-d; Supplementary Figure 12a-b). The $p$-values were 0.0038 and 0.031 for two cohorts, respectively. We also found that the MM-TCR cohort patients in the CR and response groups exhibited relatively high antigenic expression levels (Supplementary Figure 12c-d). These findings suggest that the patients with highly scored pHLA and pTCR bindings predicted by UnifyImmun benefited more from immunotherapy and yielded favourable clinical outcomes.

\section{Discussion and Conclusion}
In this study, we introduced UnifyImmun, a unified cross-attention model designed to simultaneously predict the binding specificity of peptide to both HLA and TCR molecules. We have devised a two-phase progressive training strategy through which the two tasks mutually cooperated to improve the performance of each other, by driving the encoders to capture more expressive features that enhance performance. To bolster the model's generalizability, we have incorporated virtual adversarial perturbation into the framework. When benchmarked against over ten previously published methods for pHLA and pTCR binding prediction, our method consistently outperformed them in both tasks on hold-out test sets and multiple external sets. Additionally, the cross-attention scores pinpointed the amino-acid sites crucial for peptide binding to receptors.

However, we acknowledge that our method still has some limitations. First, our model integrated the prediction tasks of pHLA and pTCR bindings into a unified framework, offering more a comprehensive evaluation of antigen immunogenicity compared to previous models that only considered individual tasks alone. However, it is important to note that the antigen-induced immune system activation involves a series of cascading biological events~\cite{chen2013oncology}, with a couple of factors influencing the antigen immunogenicity. They include endopeptidase preferences for polypeptide cleavage sites, antigen concentration, stability of pHLA complexes, and transporter protein efficiency, all of which affect the degree of immune response activation. While our model marks a significant advancement in the holistic assessment of antigen immunogenicity, it remains a high-level simplification of the actual immune response process.

Second, the currently available TCR CDR3 sequences constitute just a very small fraction of the immense TCR repertoire. This poses a significant challenge for developing method to predict pTCR binding specificity. Although our model can capture underlying patterns for antigen recognition from the TCR sequences, its capacity is still hindered by the scarcity of available data. This problem becomes particularly serious when confronted with unseen peptides in the test set. This might be the reason why our method showed only moderate performance on the real COVID-19 test set. Fortunately, the remarkable progress in single-cell transcriptome sequencing has led to a significant increase in scRNA-seq data of T cells, greatly facilitating the acquisition of CDR3 sequences. By leveraging the power of large language models (LLMs) for pretraining, we can extract more expressive and meaningful features from the massive sequences~\cite{fang2024large}. This would significantly enhance the predictive capabilities of our model to accurately assess the immunogenicity of antigens. 

Finally, the currently available training samples are actually biased to certain epitopes and their clonally expanded pairing TCRs. Compared to the vast generation space of unseen peptides, such as neoantigens and exogenous virus peptides, the number of epitopes is very limited. Also, the strategy to generate negative pTCR samples is also biased from normal protein distributions. These issues would lead to overfitting of our model, resulting in unsatisfactory performance on unseen epitopes. Therefore, it is needed to consider new strategy to generate more generic negative samples, so that we can learn a more robust model.

\section{Methods}
\subsection{Dataset} \label{sec:dataset}
In this study, we consider only the HLA class I molecules. We created a benchmark dataset \cite{Anthem,reche2005epimhc,bhasin2003mhcbn,rammensee1999syfpeithi} of pHLA bindings from over ten previous studies (for more details see Supplementary Table 1). After removal of duplicates and abnormal sequences (such as missing values or asterisk), we obtained 410,422 pHLA bindings, spanning 142 HLA alleles and 279,924 unique peptides. The frequency of amino acids in the HLA pseudo sequences and peptides bound by HLA molecules are shown in Supplementary Figure 1(a,d,f). The pHLA binding dataset was split into the training set and hold-out test set by 9:1 ratio. As a result, the training set contained 322,471 pairs, spanning 139 HLA alleles and 219,744 antigens. The independent test set contained 35,968 pairs, covering 118 HLA alleles and 33,606 antigens. We generated approximately twice the number of negative pHLA samples through two ways: \textit{random mismatching} and \textit{unbound sequence pool}. Random mismatching is done by shuffling HLA and peptide sequences and then randomly pairing them. Although this method can result in negative samples containing peptides and HLAs identical to those in the positive samples—potentially introducing negative sample bias~\cite{2023The}—the occurrence or proportion of such false negatives is minimal and can be considered negligible.
In contrast, the second way involved retrieving long protein sequences from the IEDB immunopeptidomes, which were then randomly segmented into shorter sequences to create an unbound sequence pool. From this pool, sequences were randomly extracted to pair with specific peptide to generate negative samples for each HLA allele. As a result, the negative samples generated by these two methods each comprised approximately 50\% of the total negative samples.

For each HLA allele, a portion of negative peptides were generated from the segments of the source proteins of IEDB HLA immunopeptidomes. Other negative samples were generated by shuffling the positive HLA and peptide sequences and then randomly mismatched. Although false negative samples may be generated, the possibility and proportion of such samples are very low and can be ignored.

To establish a large-scale benchmark dataset \cite{bagaev2020vdjdb,10x2019new,IEDB2018,heikkila2020human,zhang2020pird,gilson2016bindingdb,dines2020immunerace,zhang2018high,tickotsky2017mcpas} of pTCR bindings, we considered both $\alpha$ and $\beta$ chains of TCR and treated them as single CDR3 sequences, since previous studies have verified that both chains are crucial for antigen recognition \cite{hudson2023can,zhao2022tuning,carter2019single,emerson2017immunosequencing,leem2018stcrdab,mayer2023measures}. By gathering pTCR binding data from a number of previous studies (Supplementary Table 1), we created a pTCR binding dataset with 137,740 pairs, covering 1,488 unique antigens, and 128,169 unique TCR CDR3 sequences. The frequency of amino acids in the TCR CDR3 sequences and peptides bound by TCR molecules is shown in Supplementary Figure 1(b,c,e,g). We employed two strategies to generate pTCR negative samples, thereby constructing three separate pTCR binding datasets for model training and evaluation. The first strategy utilized random mismatching to generate an equal number of negative samples corresponding to the positive samples. The second strategy, termed the hybrid strategy, combined random mismatching and an unbound sequence pool, with each contributing 50\% of the total negative samples. Using the hybrid strategy, we established two other datasets, one with 1:1 positive-to-negative sample ratio and another with 1:5 positive-to-negative sample ratio. All datasets were divided into training and test sets at a 9:1 ratio to ensure robust evaluation. 

To the best of our knowledge, both the pHLA and pTCR binding datasets we built are the largest to date. 

\subsection{Sequence embedding}
The HLA pseudo sequences have a fixed length of 34 amino acids. Each amino acid is mapped to a 64-dimensional embedding via a character embedding layer. Since the order of amino acids is critical to the protein structure and function, the sine and cosine positional encoding is applied to each position. The amino-acid embedding and positional embedding are summed to obtain the sequence embedding. As a result, each HLA pseudo-sequence is represented as a 34$\times$64 matrix. 

The antigen peptides are padded to a maximum length of 15 to handle the variable input length, and then each amino acid is mapped to a 64-dimensional embedding. Similarly, the positional encoding is applied to incorporate positional information of each amino acid. After the padding and embedding steps, each peptide is represented as a 15$\times$64 matrix. 

All the TCR CDR3 sequences shorter than 34 amino acids are padded to 34, while a small portion of CDR3 sequences exceeding 34 amino acids are truncated. Next, a similar embedding process is applied to each CDR3 sequence, resulting in a 34$\times$64 representation matrix. 

\subsection{Self-attention encoder}
The encoder is based on the self-attention mechanism \cite{AttentionisAllyouNeed}, which has shown exceptional capability in extracting global dependency relationships from protein sequences \cite{madani2023large,brandes2022proteinbert,rives2021biological}. Self-attention mechanism learns the attention scores for all possible amino acid pairs within the input sequence. It computes the attention weights from the normalized dot product of query vectors $Q$ and key vectors $K$ followed by a softmax operation, and outputs the weighted sum of the value vectors $V$ by the attention scores. The operations of a self-attention layer written in matrix form are as follows:
\begin{equation}
\mathrm{Attention}(Q,K,V)=\mathrm{softmax}\left(\frac{\mathrm{QK}^T}{\sqrt{d_k}}\right)V
\end{equation}
where $d_k$ is the dimension of the vectors (chosen as 64). Taking HLA as an example, the $Q$, $K$, $V$ are all set to its 34$\times$64 embedding matrix. Subsequently, the output of the self-attention block is passed through feed-forward layers: first expanding to 512 dimensions with ReLU activation function, and then compressing to 64 for a condensed representation. The peptides and TCR encoders share the same network architecture but have independent parameters.

It is important to highlight that we introduce the mask mechanism in calculating the self-attention scores for peptides and CDR3 sequences. Specifically, for the peptides or CDR3 sequences shorter than their respective maximum lengths, we exclude non-amino-acid characters from consideration during model training. For this purpose, we assign zero attention scores corresponding to these characters, so that they do not influence the computation of attention scores. In our implementation, the encoder comprises a one-layer, one-head self-attention block. %Given a 34$\times$64 embedding matrix as input, the encoder generates an output matrix with the same dimensions.

\subsection{Cross-attention for feature fusion}
Cross-attention mechanism has been demonstrated to effectively capture the intricate relationships and global dependencies between different sequences \cite{honda2020cross,dens2023cross}. Therefore, we leverage the cross-attention mechanism to fuse the feature regarding the interactions between peptide and HLA/TCR molecules. The calculation of cross-attention scores is similar to self-attention. For the fusion of HLA and peptide feature, the HLA embedding matrix acts as the $K$ and $V$, while the peptide embedding matrix serves as the $Q$. Subsequently, the $V$ matrix is weighted by the cross-attention scores computed between $Q$ and $K$. The output of the cross-attention block passes through two feed-forward layers, by which the dimension first rises and then falls. The similar process is applied for the fusion of TCR and peptide features, where TCR embedding acts as $K$ and $V$ matrices, and peptide embedding serves as $Q$ matrix. %The cross-attention module yields the fused representations (34$\times$64) for HLA-peptide and TCR-peptide pairs, respectively.

We also employ the mask mechanism when calculating the cross-attention scores, because the mask mechanism greatly reduces the computational overhead and accelerates the model convergence. In our implementation, we adopt one-layer, one-head cross-attention mechanism, as illustrated in Figure~\ref{fig:framework}.

\subsection{Prediction of binding specificity}
To predict the bindings between peptides and HLA (or TCR) molecules, we flatten the fused matrix of HLA (or TCR)-peptide pairs outputted by the cross-attention block, resulting in a 2176-dimensional vector (aka 34×64). This vector then passes through three fully connected layers with 256, 64, and 2 nodes, utilizing the ReLU activation function. The final output is obtained via a softmax layer. We adopt cross-entropy as the loss function and use the Adam optimizer with a learning rate of 1e-3. 

The model training is conducted on a CentOS Linux 8.2.2004 (Core) system, equipped with an Intel(R) Xeon(R) Silver 4210R CPU operating at 2.40GHz, along with a GeForce RTX 4090 GPU and 128GB of memory. The model is implemented using PyTorch 2.2.1. On the large-scale benchmark dataset we built, one epoch took about 6 hours when the batch size was set to 8,192 (Supplementary Figure 2). When tested on a set with 100,000 samples, model inference finished within 10 seconds. 

\subsection{Virtual adversarial training}\label{sec:adversarial}
The virtual adversarial training \cite{miyato2017adversarial} introduces subtle perturbations within the vicinity of the sequence embedding space, rather than directly perturbing the original sequences. The perturbations are oriented toward the direction of loss gradient ascent and are typically generated under L2 norm constraints. This training strategy demands that the model not only minimizes the empirical risk but also minimizes the adversarial loss, making the model less sensitive to slight changes in the input. Formally, the adversarial loss is defined as below:
\begin{equation}
    \begin{aligned}
    L_{\mathrm{vadv}}(x,\theta)&=D\left[p(y|x,\hat{\theta}),p(y|x+r_{\mathrm{vadv}},\theta)\right]),
\\\text{ where }
r_{\mathrm{vadv}}&=\arg\max_{r;\|r\|\leq\epsilon}D\left[p(y|x_*,\hat{\theta}),p(y|x+r)\right], \end{aligned}
\end{equation}
$D$ represents the function that measures the divergence between two distributions, $p(y|x)$ denotes the probability of the model predicting label $y$ given input $x$, and $r_{\mathrm{vadv}}$ is a virtual adversarial perturbation regarding the input sample $x$. This perturbation strives to maximize the divergence between $p(y|x_*,\hat{\theta})$ and $p(y|x+r)$ by following the direction of gradient ascent.

We apply adversarial perturbations to the embeddings of all three types of sequences, so that the encoder learns to extract discriminative features. Our ablation experiments have confirmed that virtual adversarial learning indeed improves model performance, as shown in Supplementary Table 2.

\subsection{Two-phase progressive training}\label{sec:twophase}
The two-phase progressive training strategy is illustrated in Figure~\ref{fig:framework}(b). In the first phase, the model is trained exclusively on the pHLA pairs, keeping the TCR encoder and the pTCR cross-attention module fixed. This enforces the model to concentrate solely on learning the intricacy of HLA-antigen interactions. In the second phase, the model is trained exclusively using the TCR-peptide pairs, with the HLA encoder and HLA-antigen cross-attention module fixed. The two phases alternate until the model performance converged. Note that throughout the alternating training process, the antigen encoder remains continuously updated and shared between the HLA-antigen and TCR-antigen binding prediction tasks. By iteratively refining the antigen encoder parameters, the model learns to capture the essential information relevant to both HLA and TCR binding, thereby enhancing its overall predictive accuracy.

\subsection{Model ablation experiments} \label{sec:ablation}
To validate the contributions of different components, we conducted ablation experiments to assess the performance in predicting pHLA and pTCR bindings. Specifically, we evaluate the attention masking, positional encoding, and virtual adversarial perturbation independently. The performance of the ablated models for pHLA and pTCR binding prediction is outlined in Supplementary Tables 2-3, respectively.

The results reveal that the removal of any component leads to a decrease in performance. Notably, the removal of virtual adversarial perturbation has the most significant impact, resulting in at least 7\% drop in AUROC for both pHLA and pTCR binding predictions. This strongly indicates that virtual adversarial training contributes to the improvement of overall performance and generalization capabilities. Furthermore, we observed that the absence of the attention masking leads to increased computational overhead during the training process. Without the mask, additional computational resources are expended to process the padding sequences, which increases computational cost.

\section*{Declarations}

\begin{itemize}

\item \textbf{Data Availability} \\
The benchmark datasets of UnifyImmun model are available at GitHub (https://github.com/hliulab/UnifyImmun) and its Zenodo (https://doi.org/10.5281/zenodo.14282419)\cite{hui_2024_14282419}. The data resources for building the benchmark datasets are detailed in the Supplementary Table 1. The external test set and neoantigen test sets for pHLA bindings are available at: https://github.com/a96123155/TransPHLA-AOMP/tree/master/Dataset. The HPV test set was obtained from the Supplmentary data of ref\cite{bonsack2019performance}. The TCR CDR3 sequence, non-synonymous mutations, host gene expression levels and immunotherapy responses of MM-TCR cohort are available at https://github.com/riazn/bms038\_analysis. The HLA alleles, tumor antigen sequences, and immunotherapy responses of MM-HLA are available at https://www.science.org/doi/10.1126/science.aad0095. The published large cohort COVID-19 dataset is available at https://clients.adaptivebiotech.com/pub/covid-2020. The 3D crystal complex is available at PDB (https://www.rcsb.org) with accession number 3MV7.

\item \textbf{Code Availability} \\
The source codes of UnifyImmun model are available GitHub (https://github.com/hliulab/UnifyImmun) and its Zenodo (https://doi.org/10.5281/zenodo.14282419)\cite{hui_2024_14282419}. Moreover, we have developed a user-friendly web server that can be freely accessed at: http://hliulab.tech/unifylmmun/.

\item \textbf{Acknowledgements} \\
This work was supported by National Natural Science Foundation of China (No. 62072058, No. 62372229), Natural Science Foundation of Jiangsu Province (No. BK20231271).

\item \textbf{Author Contributions} \\
H.L. and  C.Y. conceptualized the idea. C.Y. implemented the model. C.Y. and X.F. collected the data and conducted the experiments. C.Y. and S.T. plotted figures. H.L. and C.Y. prepared the manuscript. H.L. revised the manuscript. H.L. supervised the research.

\item \textbf{Competing Interests} \\
The authors declare no competing interests.
% \item Ethics approval and consent to participate \\
% Not applicable.
% %\item Consent for publication

\end{itemize}

% \noindent
% If any of the sections are not relevant to your manuscript, please include the heading and write `Not applicable' for that section. 



\subsection*{Figure Captions}

% \textbf{Fig.1}: Illustrative diagram of UnifyImmun framework and two-phase training strategy, as well as the sequence frequency distributions of the benchmark datasets. (a) Architecture of UnifyImmun based on cross-attention mechanism. (b) Two-stage progressive training strategy. (c-d) Frequency of antigen sequences and TCR CDR3 sequences included in our created benchmark datasets with respect to lengths. 

% \textbf{Fig.2}: Performance evaluation on predicting peptide-HLA binding specificity. (a) Performance comparison to twelve existing methods on independent (left) and external (right) test dataset, respectively. (b) ROC curves and AUC values achieved by UnifyImmun and eight competing methods on hold-out independent test set. (c) UMAP feature visualization of peptide-HLA pairs. (d) Positive predictive value (PPV) for the top 100, top 1000, and top 5000 predicted pHLA samples. 

% \textbf{Fig.3}: Performance evaluation on predicting peptide-TCR binding specificity. (a-c) Performance comparison to four methods on independent, external, and COVID-19 test sets, respectively. (d-f) Positive predictive value (PPV) for the top 100, top 1000, and top 5000 predicted samples on independent, external, and COVID-19 test sets, respectively. (g-h) ROC curves and AUC values on independent and external test dataset, respectively. 

%\textbf{Fig.4}: Two-phase progressive training improved performance for both pHLA and pTCR binding prediction tasks. (a-b) AUROC and AUPR values increased with two-phase training rounds on pHLA independent test set. (c-d) AUROC and AUPR values increased with two-phase training rounds on the pTCR independent test set. In all the box plots, the sample size is 10 for each round, and each point represents the result of a technical replicate by randomly data partition. The horizontal line within each box represents the median, the box boundaries denote the interquartile range (Q25 to Q75), and whiskers extend to the most extreme data point no more than 1.5 times the interquartile range. The one-sided $F$-test was performed to assess the statistical differences between the two groups connected by a line, with the corresponding p-values displayed above the respective lines.

%\textbf{Fig.5}: Heatmaps generated from cross-attention scores and integrated gradients. (a-b) Heatmaps of cross-attention scores and integrated gradients of the amino-acid type at each position of 9-mer peptide binding to HLA molecules. (c,f) Accumulative attention scores across peptide length of each amino-acid type of peptide binding to HLA and TCR molecules, respectively. (d-e) Heatmaps of cross-attention scores and integrated gradients of the amino-acid type at each position of 9-mer peptide binding to TCR molecules. (g) Heatmaps of cross-attention scores for top five HLA alleles with most 9-mer binding peptides. (h-i) Attention score-based heatmap and 3D structure for TCR complex with HLA-B35:01/HPVG (PDB ID: 3MV7).

%\textbf{Fig.6}: Correlation between UnifyImmun predicted binding scores and immunotherapy response and clinical outcomes on two clinical cohorts. (a) Violin plots of predicted pHLA binding scores regarding the different immunotherapy response groups of MM-HLA cohort with CR (n=632), PR (n=15,916), SD (n=15,562), and PD (n=68,076), respectively. (b) Violin plots of predicted pTCR binding scores regarding the different immunotherapy response groups of MM-TCR cohort with CR (n=10,000), PR (n=25,000), SD (n=45,000), and PD (n=60,000), respectively. For the boxplots within the violin plot, the horizontal line within each box represents the median, the box boundaries denote the interquartile range (Q25 to Q75), and whiskers extend to the most extreme data point no more than 1.5 times the interquartile range. The one-sided $F$-test was performed to assess the statistical differences between the two groups connected by a line, with the corresponding p-values displayed above the respective lines. (c-d) Survival curves between stratified patient groups with high- and low-confidence antigen binding specificity on MM-HLA and MM-TCR cohorts, respectively.

\bibliography{reference}% common bib file

\begin{thebibliography}{10}
\expandafter\ifx\csname url\endcsname\relax
  \def\url#1{\burl{#1}}\fi
\expandafter\ifx\csname urlprefix\endcsname\relax\def\urlprefix{URL }\fi
\providecommand{\bibinfo}[2]{#2}
\providecommand{\eprint}[2][]{\url{#2}}
\providecommand{\doi}[1]{\url{https://doi.org/#1}}
\bibcommenthead

\bibitem{schumacher2015neoantigens}
\bibinfo{author}{Schumacher, T.~N.} \& \bibinfo{author}{Schreiber, R.~D.}
\newblock \bibinfo{title}{Neoantigens in cancer immunotherapy}.
\newblock \emph{\bibinfo{journal}{Science}} \textbf{\bibinfo{volume}{348}}, \bibinfo{pages}{69--74} (\bibinfo{year}{2015}).

\bibitem{glanville2017identifying}
\bibinfo{author}{Glanville, J.} \emph{et~al.}
\newblock \bibinfo{title}{Identifying specificity groups in the t cell receptor repertoire}.
\newblock \emph{\bibinfo{journal}{Nature}} \textbf{\bibinfo{volume}{547}}, \bibinfo{pages}{94--98} (\bibinfo{year}{2017}).

\bibitem{zhang2019combination}
\bibinfo{author}{Zhang, J.} \emph{et~al.}
\newblock \bibinfo{title}{The combination of neoantigen quality and t lymphocyte infiltrates identifies glioblastomas with the longest survival}.
\newblock \emph{\bibinfo{journal}{Communications biology}} \textbf{\bibinfo{volume}{2}}, \bibinfo{pages}{135} (\bibinfo{year}{2019}).

\bibitem{yang2023antigen}
\bibinfo{author}{Yang, K.}, \bibinfo{author}{Halima, A.} \& \bibinfo{author}{Chan, T.~A.}
\newblock \bibinfo{title}{Antigen presentation in cancer—mechanisms and clinical implications for immunotherapy}.
\newblock \emph{\bibinfo{journal}{Nature Reviews Clinical Oncology}} \textbf{\bibinfo{volume}{20}}, \bibinfo{pages}{604--623} (\bibinfo{year}{2023}).

\bibitem{barry2002cytotoxic}
\bibinfo{author}{Barry, M.} \& \bibinfo{author}{Bleackley, R.~C.}
\newblock \bibinfo{title}{Cytotoxic t lymphocytes: all roads lead to death}.
\newblock \emph{\bibinfo{journal}{Nature Reviews Immunology}} \textbf{\bibinfo{volume}{2}}, \bibinfo{pages}{401--409} (\bibinfo{year}{2002}).

\bibitem{raskov2021cytotoxic}
\bibinfo{author}{Raskov, H.}, \bibinfo{author}{Orhan, A.}, \bibinfo{author}{Christensen, J.~P.} \& \bibinfo{author}{G{\"o}genur, I.}
\newblock \bibinfo{title}{Cytotoxic cd8+ t cells in cancer and cancer immunotherapy}.
\newblock \emph{\bibinfo{journal}{British journal of cancer}} \textbf{\bibinfo{volume}{124}}, \bibinfo{pages}{359--367} (\bibinfo{year}{2021}).

\bibitem{weigelin2021cytotoxic}
\bibinfo{author}{Weigelin, B.} \emph{et~al.}
\newblock \bibinfo{title}{Cytotoxic t cells are able to efficiently eliminate cancer cells by additive cytotoxicity}.
\newblock \emph{\bibinfo{journal}{Nature communications}} \textbf{\bibinfo{volume}{12}}, \bibinfo{pages}{5217} (\bibinfo{year}{2021}).

\bibitem{rowen1996complete}
\bibinfo{author}{Rowen, L.}, \bibinfo{author}{Koop, B.~F.} \& \bibinfo{author}{Hood, L.}
\newblock \bibinfo{title}{The complete 685-kilobase dna sequence of the human $\beta$ t cell receptor locus}.
\newblock \emph{\bibinfo{journal}{Science}} \textbf{\bibinfo{volume}{272}}, \bibinfo{pages}{1755--1762} (\bibinfo{year}{1996}).

\bibitem{fowell2021spatio}
\bibinfo{author}{Fowell, D.~J.} \& \bibinfo{author}{Kim, M.}
\newblock \bibinfo{title}{The spatio-temporal control of effector t cell migration}.
\newblock \emph{\bibinfo{journal}{Nature Reviews Immunology}} \textbf{\bibinfo{volume}{21}}, \bibinfo{pages}{582--596} (\bibinfo{year}{2021}).

\bibitem{lim2020tumor}
\bibinfo{author}{Lim, A.~R.}, \bibinfo{author}{Rathmell, W.~K.} \& \bibinfo{author}{Rathmell, J.~C.}
\newblock \bibinfo{title}{The tumor microenvironment as a metabolic barrier to effector t cells and immunotherapy}.
\newblock \emph{\bibinfo{journal}{Elife}} \textbf{\bibinfo{volume}{9}}, \bibinfo{pages}{e55185} (\bibinfo{year}{2020}).

\bibitem{chen2013oncology}
\bibinfo{author}{Chen, D.~S.} \& \bibinfo{author}{Mellman, I.}
\newblock \bibinfo{title}{Oncology meets immunology: the cancer-immunity cycle}.
\newblock \emph{\bibinfo{journal}{immunity}} \textbf{\bibinfo{volume}{39}}, \bibinfo{pages}{1--10} (\bibinfo{year}{2013}).

\bibitem{yewdell1999immunodominance}
\bibinfo{author}{Yewdell, J.~W.} \& \bibinfo{author}{Bennink, J.~R.}
\newblock \bibinfo{title}{Immunodominance in major histocompatibility complex class i--restricted t lymphocyte responses}.
\newblock \emph{\bibinfo{journal}{Annual review of immunology}} \textbf{\bibinfo{volume}{17}}, \bibinfo{pages}{51--88} (\bibinfo{year}{1999}).

\bibitem{dunn2004three}
\bibinfo{author}{Dunn, G.~P.}, \bibinfo{author}{Old, L.~J.} \& \bibinfo{author}{Schreiber, R.~D.}
\newblock \bibinfo{title}{The three es of cancer immunoediting}.
\newblock \emph{\bibinfo{journal}{Annu. Rev. Immunol.}} \textbf{\bibinfo{volume}{22}}, \bibinfo{pages}{329--360} (\bibinfo{year}{2004}).

\bibitem{trowsdale2005hla}
\bibinfo{author}{Trowsdale, J.}
\newblock \bibinfo{title}{Hla genomics in the third millennium}.
\newblock \emph{\bibinfo{journal}{Current opinion in Immunology}} \textbf{\bibinfo{volume}{17}}, \bibinfo{pages}{498--504} (\bibinfo{year}{2005}).

\bibitem{huppa2010tcr}
\bibinfo{author}{Huppa, J.~B.} \emph{et~al.}
\newblock \bibinfo{title}{Tcr--peptide--mhc interactions in situ show accelerated kinetics and increased affinity}.
\newblock \emph{\bibinfo{journal}{Nature}} \textbf{\bibinfo{volume}{463}}, \bibinfo{pages}{963--967} (\bibinfo{year}{2010}).

\bibitem{nikolich2004many}
\bibinfo{author}{Nikolich-{\v{Z}}ugich, J.}, \bibinfo{author}{Slifka, M.~K.} \& \bibinfo{author}{Messaoudi, I.}
\newblock \bibinfo{title}{The many important facets of t-cell repertoire diversity}.
\newblock \emph{\bibinfo{journal}{Nature Reviews Immunology}} \textbf{\bibinfo{volume}{4}}, \bibinfo{pages}{123--132} (\bibinfo{year}{2004}).

\bibitem{zhang2016direct}
\bibinfo{author}{Zhang, S.-Q.} \emph{et~al.}
\newblock \bibinfo{title}{Direct measurement of t cell receptor affinity and sequence from na{\"\i}ve antiviral t cells}.
\newblock \emph{\bibinfo{journal}{Science translational medicine}} \textbf{\bibinfo{volume}{8}}, \bibinfo{pages}{341ra77--341ra77} (\bibinfo{year}{2016}).

\bibitem{davis1988t}
\bibinfo{author}{Davis, M.~M.} \& \bibinfo{author}{Bjorkman, P.~J.}
\newblock \bibinfo{title}{T-cell antigen receptor genes and t-cell recognition}.
\newblock \emph{\bibinfo{journal}{Nature}} \textbf{\bibinfo{volume}{334}}, \bibinfo{pages}{395--402} (\bibinfo{year}{1988}).

\bibitem{krogsgaard2005t}
\bibinfo{author}{Krogsgaard, M.} \& \bibinfo{author}{Davis, M.~M.}
\newblock \bibinfo{title}{How t cells' see'antigen}.
\newblock \emph{\bibinfo{journal}{Nature immunology}} \textbf{\bibinfo{volume}{6}}, \bibinfo{pages}{239--245} (\bibinfo{year}{2005}).

\bibitem{chowell2019evolutionary}
\bibinfo{author}{Chowell, D.} \emph{et~al.}
\newblock \bibinfo{title}{Evolutionary divergence of hla class i genotype impacts efficacy of cancer immunotherapy}.
\newblock \emph{\bibinfo{journal}{Nature medicine}} \textbf{\bibinfo{volume}{25}}, \bibinfo{pages}{1715--1720} (\bibinfo{year}{2019}).

\bibitem{krishna2020genetic}
\bibinfo{author}{Krishna, C.}, \bibinfo{author}{Chowell, D.}, \bibinfo{author}{G{\"o}nen, M.}, \bibinfo{author}{Elhanati, Y.} \& \bibinfo{author}{Chan, T.~A.}
\newblock \bibinfo{title}{Genetic and environmental determinants of human tcr repertoire diversity}.
\newblock \emph{\bibinfo{journal}{Immunity \& Ageing}} \textbf{\bibinfo{volume}{17}}, \bibinfo{pages}{1--7} (\bibinfo{year}{2020}).

\bibitem{purcell2019mass}
\bibinfo{author}{Purcell, A.~W.}, \bibinfo{author}{Ramarathinam, S.~H.} \& \bibinfo{author}{Ternette, N.}
\newblock \bibinfo{title}{Mass spectrometry--based identification of mhc-bound peptides for immunopeptidomics}.
\newblock \emph{\bibinfo{journal}{Nature protocols}} \textbf{\bibinfo{volume}{14}}, \bibinfo{pages}{1687--1707} (\bibinfo{year}{2019}).

\bibitem{zhang2018high}
\bibinfo{author}{Zhang, S.-Q.} \emph{et~al.}
\newblock \bibinfo{title}{High-throughput determination of the antigen specificities of t cell receptors in single cells}.
\newblock \emph{\bibinfo{journal}{Nature biotechnology}} \textbf{\bibinfo{volume}{36}}, \bibinfo{pages}{1156--1159} (\bibinfo{year}{2018}).

\bibitem{kula2019t}
\bibinfo{author}{Kula, T.} \emph{et~al.}
\newblock \bibinfo{title}{T-scan: a genome-wide method for the systematic discovery of t cell epitopes}.
\newblock \emph{\bibinfo{journal}{Cell}} \textbf{\bibinfo{volume}{178}}, \bibinfo{pages}{1016--1028} (\bibinfo{year}{2019}).

\bibitem{hudson2023can}
\bibinfo{author}{Hudson, D.}, \bibinfo{author}{Fernandes, R.~A.}, \bibinfo{author}{Basham, M.}, \bibinfo{author}{Ogg, G.} \& \bibinfo{author}{Koohy, H.}
\newblock \bibinfo{title}{Can we predict t cell specificity with digital biology and machine learning?}
\newblock \emph{\bibinfo{journal}{Nature Reviews Immunology}} \textbf{\bibinfo{volume}{23}}, \bibinfo{pages}{511--521} (\bibinfo{year}{2023}).

\bibitem{TransPHLA2022}
\bibinfo{author}{Chu, Y.} \emph{et~al.}
\newblock \bibinfo{title}{A transformer-based model to predict peptide--hla class i binding and optimize mutated peptides for vaccine design}.
\newblock \emph{\bibinfo{journal}{Nature Machine Intelligence}} \textbf{\bibinfo{volume}{4}}, \bibinfo{pages}{300--311} (\bibinfo{year}{2022}).

\bibitem{2017MHCflurry}
\bibinfo{author}{O'Donnell, T.}, \bibinfo{author}{Rubinsteyn, A.}, \bibinfo{author}{Bonsack, M.}, \bibinfo{author}{Riemer, A.} \& \bibinfo{author}{Hammerbacher, J.}
\newblock \bibinfo{title}{Mhcflurry: open-source class i mhc binding affinity prediction}.
\newblock \emph{\bibinfo{journal}{Cold Spring Harbor Laboratory}}  (\bibinfo{year}{2017}).

\bibitem{NetMHCpan}
\bibinfo{author}{Birkir, R.}, \bibinfo{author}{Bruno, A.}, \bibinfo{author}{Sinu, P.}, \bibinfo{author}{Bjoern, P.} \& \bibinfo{author}{Morten, N.}
\newblock \bibinfo{title}{Netmhcpan-4.1 and netmhciipan-4.0: improved predictions of mhc antigen presentation by concurrent motif deconvolution and integration of ms mhc eluted ligand data}.
\newblock \emph{\bibinfo{journal}{Nucleic Acids Research}} .

\bibitem{2019DeepLigand}
\bibinfo{author}{Haoyang, Z.} \& \bibinfo{author}{Gifford, D.~K.}
\newblock \bibinfo{title}{Deepligand: accurate prediction of mhc class i ligands using peptide embedding}.
\newblock \emph{\bibinfo{journal}{Bioinformatics}} \bibinfo{pages}{i278--i283} (\bibinfo{year}{2019}).

\bibitem{2021BERTMHC}
\bibinfo{author}{Jun, C.}, \bibinfo{author}{Kadre, B.}, \bibinfo{author}{Karola, R.} \& \bibinfo{author}{Brandon, M.}
\newblock \bibinfo{title}{Bertmhc: improved mhc–peptide class ii interaction prediction with transformer and multiple instance learning}.
\newblock \emph{\bibinfo{journal}{Bioinformatics}}  (\bibinfo{year}{2021}).

\bibitem{PanPep2023}
\bibinfo{author}{Gao, Y.} \emph{et~al.}
\newblock \bibinfo{title}{Pan-peptide meta learning for t-cell receptor--antigen binding recognition}.
\newblock \emph{\bibinfo{journal}{Nature Machine Intelligence}} \textbf{\bibinfo{volume}{5}}, \bibinfo{pages}{236--249} (\bibinfo{year}{2023}).

\bibitem{pMTnet2021}
\bibinfo{author}{Lu, T.} \emph{et~al.}
\newblock \bibinfo{title}{Deep learning-based prediction of the t cell receptor--antigen binding specificity}.
\newblock \emph{\bibinfo{journal}{Nature machine intelligence}} \textbf{\bibinfo{volume}{3}}, \bibinfo{pages}{864--875} (\bibinfo{year}{2021}).

\bibitem{2021DLpTCR}
\bibinfo{author}{Zhaochun, X.} \emph{et~al.}
\newblock \bibinfo{title}{Dlptcr: an ensemble deep learning framework for predicting immunogenic peptide recognized by t cell receptor}.
\newblock \emph{\bibinfo{journal}{Briefings in Bioinformatics}}  (\bibinfo{year}{2021}).

\bibitem{ERGO22021}
\bibinfo{author}{Springer, I.}, \bibinfo{author}{Tickotsky, N.} \& \bibinfo{author}{Louzoun, Y.}
\newblock \bibinfo{title}{Contribution of t cell receptor alpha and beta cdr3, mhc typing, v and j genes to peptide binding prediction}.
\newblock \emph{\bibinfo{journal}{Frontiers in Immunology}} \textbf{\bibinfo{volume}{12}}, \bibinfo{pages}{664514} (\bibinfo{year}{2021}).

\bibitem{2021TITAN}
\bibinfo{author}{Weber, A.}, \bibinfo{author}{Born, J.} \& \bibinfo{author}{Martínez, M.~R.}
\newblock \bibinfo{title}{Titan: T-cell receptor specificity prediction with bimodal attention networks}.
\newblock \emph{\bibinfo{journal}{Bioinformatics}} \textbf{\bibinfo{volume}{37}}, \bibinfo{pages}{i237--i244} (\bibinfo{year}{2021}).

\bibitem{fang2022attention}
\bibinfo{author}{Fang, Y.}, \bibinfo{author}{Liu, X.} \& \bibinfo{author}{Liu, H.}
\newblock \bibinfo{title}{Attention-aware contrastive learning for predicting t cell receptor--antigen binding specificity}.
\newblock \emph{\bibinfo{journal}{Briefings in Bioinformatics}} \textbf{\bibinfo{volume}{23}}, \bibinfo{pages}{bbac378} (\bibinfo{year}{2022}).

\bibitem{meysman2023benchmarking}
\bibinfo{author}{Meysman, P.} \emph{et~al.}
\newblock \bibinfo{title}{Benchmarking solutions to the t-cell receptor epitope prediction problem: Immrep22 workshop report}.
\newblock \emph{\bibinfo{journal}{ImmunoInformatics}} \textbf{\bibinfo{volume}{9}}, \bibinfo{pages}{100024} (\bibinfo{year}{2023}).

\bibitem{ANN}
\bibinfo{author}{Massimo, A.} \& \bibinfo{author}{Morten, N.}
\newblock \bibinfo{title}{Gapped sequence alignment using artificial neural networks: application to the mhc class i system}.
\newblock \emph{\bibinfo{journal}{Bioinformatics}} \bibinfo{pages}{511}.

\bibitem{PickPocket}
\bibinfo{author}{Zhang, H.}, \bibinfo{author}{Lund, O.} \& \bibinfo{author}{Nielsen, M.}
\newblock \bibinfo{title}{The pickpocket method for predicting binding specificities for receptors based on receptor pocket similarities}.
\newblock \emph{\bibinfo{journal}{Bioinformatics}}  (\bibinfo{year}{2009}).

\bibitem{SMMPMBEC}
\bibinfo{author}{Kim, Y.}, \bibinfo{author}{Sidney, J.}, \bibinfo{author}{Pinilla, C.}, \bibinfo{author}{Sette, A.} \& \bibinfo{author}{Peters, B.}
\newblock \bibinfo{title}{Derivation of an amino acid similarity matrix for peptide:mhc binding and its application as a bayesian prior}.
\newblock \emph{\bibinfo{journal}{BMC Bioinformatics}} \textbf{\bibinfo{volume}{10}} (\bibinfo{year}{2009}).

\bibitem{SMM}
\bibinfo{author}{Peters, B.} \& \bibinfo{author}{Sette, A.}
\newblock \bibinfo{title}{Generating quantitative models describing the sequence specificity of biological processes with the stabilized matrix method}.
\newblock \emph{\bibinfo{journal}{BMC Bioinformatics}} \textbf{\bibinfo{volume}{6}}, \bibinfo{pages}{132 -- 132} (\bibinfo{year}{2005}).

\bibitem{2012NetMHCcons}
\bibinfo{author}{Karosiene, E.}, \bibinfo{author}{Lundegaard, C.}, \bibinfo{author}{Lund, O.} \& \bibinfo{author}{Nielsen, M.}
\newblock \bibinfo{title}{Netmhccons: a consensus method for the major histocompatibility complex class i predictions}.
\newblock \emph{\bibinfo{journal}{Immunogenetics}} \textbf{\bibinfo{volume}{64}}, \bibinfo{pages}{177--186} (\bibinfo{year}{2012}).

\bibitem{Rasmussen2016Pan}
\bibinfo{author}{Rasmussen} \emph{et~al.}
\newblock \bibinfo{title}{Pan-specific prediction of peptide-mhc class i complex stability, a correlate of t cell immunogenicity.}
\newblock \emph{\bibinfo{journal}{Journal of Immunology}}  (\bibinfo{year}{2016}).

\bibitem{2006A}
\bibinfo{author}{Moutaftsi, M.} \emph{et~al.}
\newblock \bibinfo{title}{A consensus epitope prediction approach identifies the breadth of murine t(cd8+)-cell responses to vaccinia virus.}
\newblock \emph{\bibinfo{journal}{Nature Publishing Group}}  (\bibinfo{year}{2006}).

\bibitem{ACME}
\bibinfo{author}{Yan, H.} \emph{et~al.}
\newblock \bibinfo{title}{Acme: pan-specific peptide–mhc class i binding prediction through attention-based deep neural networks}.
\newblock \emph{\bibinfo{journal}{Bioinformatics}} \bibinfo{pages}{23}.

\bibitem{2021Deep}
\bibinfo{author}{Jin, J.} \emph{et~al.}
\newblock \bibinfo{title}{Deep learning pan-specific model for interpretable mhc-i peptide binding prediction with improved attention mechanism}.
\newblock \emph{\bibinfo{journal}{Proteins}} \textbf{\bibinfo{volume}{89}}, \bibinfo{pages}{866--883} (\bibinfo{year}{2021}).

\bibitem{Anthem}
\bibinfo{author}{Mei, S.} \emph{et~al.}
\newblock \bibinfo{title}{Anthem: a user customised tool for fast and accurate prediction of binding between peptides and hla class i molecules}.
\newblock \emph{\bibinfo{journal}{Briefings in Bioinformatics}} \textbf{\bibinfo{volume}{22}}, \bibinfo{pages}{bbaa415} (\bibinfo{year}{2021}).

\bibitem{bonsack2019performance}
\bibinfo{author}{Bonsack, M.} \emph{et~al.}
\newblock \bibinfo{title}{Performance evaluation of mhc class-i binding prediction tools based on an experimentally validated mhc--peptide binding data set}.
\newblock \emph{\bibinfo{journal}{Cancer immunology research}} \textbf{\bibinfo{volume}{7}}, \bibinfo{pages}{719--736} (\bibinfo{year}{2019}).

\bibitem{wells2020key}
\bibinfo{author}{Wells, D.~K.} \emph{et~al.}
\newblock \bibinfo{title}{Key parameters of tumor epitope immunogenicity revealed through a consortium approach improve neoantigen prediction}.
\newblock \emph{\bibinfo{journal}{Cell}} \textbf{\bibinfo{volume}{183}}, \bibinfo{pages}{818--834} (\bibinfo{year}{2020}).

\bibitem{wang2020ineo}
\bibinfo{author}{Wang, G.} \emph{et~al.}
\newblock \bibinfo{title}{Ineo-epp: a novel t-cell hla class-i immunogenicity or neoantigenic epitope prediction method based on sequence-related amino acid features}.
\newblock \emph{\bibinfo{journal}{BioMed research international}} \textbf{\bibinfo{volume}{2020}} (\bibinfo{year}{2020}).

\bibitem{parker1992sequence}
\bibinfo{author}{Parker, K.~C.} \emph{et~al.}
\newblock \bibinfo{title}{Sequence motifs important for peptide binding to the human mhc class i molecule, hla-a2.}
\newblock \emph{\bibinfo{journal}{Journal of immunology (Baltimore, Md.: 1950)}} \textbf{\bibinfo{volume}{149}}, \bibinfo{pages}{3580--3587} (\bibinfo{year}{1992}).

\bibitem{dibrino1993hla}
\bibinfo{author}{Dibrino, M.} \emph{et~al.}
\newblock \bibinfo{title}{Hla-a1 and hla-a3 t cell epitopes derived from influenza virus proteins predicted from peptide binding motifs.}
\newblock \emph{\bibinfo{journal}{Journal of immunology (Baltimore, Md.: 1950)}} \textbf{\bibinfo{volume}{151}}, \bibinfo{pages}{5930--5935} (\bibinfo{year}{1993}).

\bibitem{dibrino1994endogenous}
\bibinfo{author}{DiBrino, M.} \emph{et~al.}
\newblock \bibinfo{title}{Endogenous peptides with distinct amino acid anchor residue motifs bind to hla-a1 and hla-b8.}
\newblock \emph{\bibinfo{journal}{Journal of immunology (Baltimore, Md.: 1950)}} \textbf{\bibinfo{volume}{152}}, \bibinfo{pages}{620--631} (\bibinfo{year}{1994}).

\bibitem{dibrino1994hla}
\bibinfo{author}{DiBrino, M.} \emph{et~al.}
\newblock \bibinfo{title}{The hla-b14 peptide binding site can accommodate peptides with different combinations of anchor residues.}
\newblock \emph{\bibinfo{journal}{Journal of Biological Chemistry}} \textbf{\bibinfo{volume}{269}}, \bibinfo{pages}{32426--32434} (\bibinfo{year}{1994}).

\bibitem{parker1994pocket}
\bibinfo{author}{Parker, K.~C.}, \bibinfo{author}{Biddison, W.~E.} \& \bibinfo{author}{Coligan, J.~E.}
\newblock \bibinfo{title}{Pocket mutations of hla-b27 show that anchor residues act cumulatively to stabilize peptide binding}.
\newblock \emph{\bibinfo{journal}{Biochemistry}} \textbf{\bibinfo{volume}{33}}, \bibinfo{pages}{7736--7743} (\bibinfo{year}{1994}).

\bibitem{dibrino1995identification}
\bibinfo{author}{DiBrino, M.} \emph{et~al.}
\newblock \bibinfo{title}{Identification of the peptide binding motif for hla-b44, one of the most common hla-b alleles in the caucasian population}.
\newblock \emph{\bibinfo{journal}{Biochemistry}} \textbf{\bibinfo{volume}{34}}, \bibinfo{pages}{10130--10138} (\bibinfo{year}{1995}).

\bibitem{nolan2020large}
\bibinfo{author}{Nolan, S.} \emph{et~al.}
\newblock \bibinfo{title}{A large-scale database of t-cell receptor beta (tcr$\beta$) sequences and binding associations from natural and synthetic exposure to sars-cov-2}.
\newblock \emph{\bibinfo{journal}{Research square}}  (\bibinfo{year}{2020}).

\bibitem{van2015genomic}
\bibinfo{author}{Van~Allen, E.~M.} \emph{et~al.}
\newblock \bibinfo{title}{Genomic correlates of response to ctla-4 blockade in metastatic melanoma}.
\newblock \emph{\bibinfo{journal}{Science}} \textbf{\bibinfo{volume}{350}}, \bibinfo{pages}{207--211} (\bibinfo{year}{2015}).

\bibitem{RIAZ2017934}
\bibinfo{title}{Tumor and microenvironment evolution during immunotherapy with nivolumab}.
\newblock \emph{\bibinfo{journal}{Cell}} \textbf{\bibinfo{volume}{171}}, \bibinfo{pages}{934--949.e16} (\bibinfo{year}{2017}).

\bibitem{fang2024large}
\bibinfo{author}{Fang, X.}, \bibinfo{author}{Yu, C.}, \bibinfo{author}{Tian, S.} \& \bibinfo{author}{Liu, H.}
\newblock \bibinfo{title}{A large language model for predicting t cell receptor-antigen binding specificity}.
\newblock \emph{\bibinfo{journal}{arXiv preprint arXiv:2406.16995}}  (\bibinfo{year}{2024}).

\bibitem{reche2005epimhc}
\bibinfo{author}{Reche, P.~A.}, \bibinfo{author}{Zhang, H.}, \bibinfo{author}{Glutting, J.-P.} \& \bibinfo{author}{Reinherz, E.~L.}
\newblock \bibinfo{title}{Epimhc: a curated database of mhc-binding peptides for customized computational vaccinology}.
\newblock \emph{\bibinfo{journal}{Bioinformatics}} \textbf{\bibinfo{volume}{21}}, \bibinfo{pages}{2140--2141} (\bibinfo{year}{2005}).

\bibitem{bhasin2003mhcbn}
\bibinfo{author}{Bhasin, M.}, \bibinfo{author}{Singh, H.} \& \bibinfo{author}{Raghava, G. P.~S.}
\newblock \bibinfo{title}{Mhcbn: a comprehensive database of mhc binding and non-binding peptides}.
\newblock \emph{\bibinfo{journal}{Bioinformatics}} \textbf{\bibinfo{volume}{19}}, \bibinfo{pages}{665--666} (\bibinfo{year}{2003}).

\bibitem{rammensee1999syfpeithi}
\bibinfo{author}{Rammensee, H.-G.}, \bibinfo{author}{Bachmann, J.}, \bibinfo{author}{Emmerich, N. P.~N.}, \bibinfo{author}{Bachor, O.~A.} \& \bibinfo{author}{Stevanovi{\'c}, S.}
\newblock \bibinfo{title}{Syfpeithi: database for mhc ligands and peptide motifs}.
\newblock \emph{\bibinfo{journal}{Immunogenetics}} \textbf{\bibinfo{volume}{50}}, \bibinfo{pages}{213--219} (\bibinfo{year}{1999}).

\bibitem{2023The}
\bibinfo{author}{Dens, C.}, \bibinfo{author}{Laukens, K.}, \bibinfo{author}{Bittremieux, W.} \& \bibinfo{author}{Meysman, P.}
\newblock \bibinfo{title}{The pitfalls of negative data bias for the t-cell epitope specificity challenge}.
\newblock \emph{\bibinfo{journal}{Nature Machine Intelligence}} \textbf{\bibinfo{volume}{5}}, \bibinfo{pages}{1060--1062} (\bibinfo{year}{2023}).

\bibitem{bagaev2020vdjdb}
\bibinfo{author}{Bagaev, D.~V.} \emph{et~al.}
\newblock \bibinfo{title}{Vdjdb in 2019: database extension, new analysis infrastructure and a t-cell receptor motif compendium}.
\newblock \emph{\bibinfo{journal}{Nucleic Acids Research}} \textbf{\bibinfo{volume}{48}}, \bibinfo{pages}{D1057--D1062} (\bibinfo{year}{2020}).

\bibitem{10x2019new}
\bibinfo{author}{10x Genomics}.
\newblock \bibinfo{title}{A new way of exploring immunity--linking highly multiplexed antigen recognition to immune repertoire and phenotype}.
\newblock \emph{\bibinfo{journal}{Tech. rep}}  (\bibinfo{year}{2019}).

\bibitem{IEDB2018}
\bibinfo{author}{Vita, R.} \emph{et~al.}
\newblock \bibinfo{title}{The immune epitope database (iedb): 2018 update}.
\newblock \emph{\bibinfo{journal}{Nucleic acids research}} \textbf{\bibinfo{volume}{47}}, \bibinfo{pages}{D339--D343} (\bibinfo{year}{2019}).

\bibitem{heikkila2020human}
\bibinfo{author}{Heikkil{\"a}, N.} \emph{et~al.}
\newblock \bibinfo{title}{Human thymic t cell repertoire is imprinted with strong convergence to shared sequences}.
\newblock \emph{\bibinfo{journal}{Molecular Immunology}} \textbf{\bibinfo{volume}{127}}, \bibinfo{pages}{112--123} (\bibinfo{year}{2020}).

\bibitem{zhang2020pird}
\bibinfo{author}{Zhang, W.} \emph{et~al.}
\newblock \bibinfo{title}{Pird: pan immune repertoire database}.
\newblock \emph{\bibinfo{journal}{Bioinformatics}} \textbf{\bibinfo{volume}{36}}, \bibinfo{pages}{897--903} (\bibinfo{year}{2020}).

\bibitem{gilson2016bindingdb}
\bibinfo{author}{Gilson, M.~K.} \emph{et~al.}
\newblock \bibinfo{title}{Bindingdb in 2015: a public database for medicinal chemistry, computational chemistry and systems pharmacology}.
\newblock \emph{\bibinfo{journal}{Nucleic acids research}} \textbf{\bibinfo{volume}{44}}, \bibinfo{pages}{D1045--D1053} (\bibinfo{year}{2016}).

\bibitem{dines2020immunerace}
\bibinfo{author}{Dines, J.~N.} \emph{et~al.}
\newblock \bibinfo{title}{The immunerace study: a prospective multicohort study of immune response action to covid-19 events with the immunecode™ open access database}.
\newblock \emph{\bibinfo{journal}{medRxiv}} \bibinfo{pages}{2020--08} (\bibinfo{year}{2020}).

\bibitem{tickotsky2017mcpas}
\bibinfo{author}{Tickotsky, N.}, \bibinfo{author}{Sagiv, T.}, \bibinfo{author}{Prilusky, J.}, \bibinfo{author}{Shifrut, E.} \& \bibinfo{author}{Friedman, N.}
\newblock \bibinfo{title}{Mcpas-tcr: a manually curated catalogue of pathology-associated t cell receptor sequences}.
\newblock \emph{\bibinfo{journal}{Bioinformatics}} \textbf{\bibinfo{volume}{33}}, \bibinfo{pages}{2924--2929} (\bibinfo{year}{2017}).

\bibitem{zhao2022tuning}
\bibinfo{author}{Zhao, X.} \emph{et~al.}
\newblock \bibinfo{title}{Tuning t cell receptor sensitivity through catch bond engineering}.
\newblock \emph{\bibinfo{journal}{Science}} \textbf{\bibinfo{volume}{376}}, \bibinfo{pages}{eabl5282} (\bibinfo{year}{2022}).

\bibitem{carter2019single}
\bibinfo{author}{Carter, J.~A.} \emph{et~al.}
\newblock \bibinfo{title}{Single t cell sequencing demonstrates the functional role of $\alpha$$\beta$ tcr pairing in cell lineage and antigen specificity}.
\newblock \emph{\bibinfo{journal}{Frontiers in immunology}} \textbf{\bibinfo{volume}{10}}, \bibinfo{pages}{1516} (\bibinfo{year}{2019}).

\bibitem{emerson2017immunosequencing}
\bibinfo{author}{Emerson, R.~O.} \emph{et~al.}
\newblock \bibinfo{title}{Immunosequencing identifies signatures of cytomegalovirus exposure history and hla-mediated effects on the t cell repertoire}.
\newblock \emph{\bibinfo{journal}{Nature genetics}} \textbf{\bibinfo{volume}{49}}, \bibinfo{pages}{659--665} (\bibinfo{year}{2017}).

\bibitem{leem2018stcrdab}
\bibinfo{author}{Leem, J.}, \bibinfo{author}{de~Oliveira, S. H.~P.}, \bibinfo{author}{Krawczyk, K.} \& \bibinfo{author}{Deane, C.~M.}
\newblock \bibinfo{title}{Stcrdab: the structural t-cell receptor database}.
\newblock \emph{\bibinfo{journal}{Nucleic acids research}} \textbf{\bibinfo{volume}{46}}, \bibinfo{pages}{D406--D412} (\bibinfo{year}{2018}).

\bibitem{mayer2023measures}
\bibinfo{author}{Mayer, A.} \& \bibinfo{author}{Callan~Jr, C.~G.}
\newblock \bibinfo{title}{Measures of epitope binding degeneracy from t cell receptor repertoires}.
\newblock \emph{\bibinfo{journal}{Proceedings of the National Academy of Sciences}} \textbf{\bibinfo{volume}{120}}, \bibinfo{pages}{e2213264120} (\bibinfo{year}{2023}).

\bibitem{AttentionisAllyouNeed}
\bibinfo{author}{Vaswani, A.} \emph{et~al.}
\newblock \bibinfo{editor}{Guyon, I.} \emph{et~al.} (eds) \emph{\bibinfo{title}{Attention is all you need}}.
\newblock (eds \bibinfo{editor}{Guyon, I.} \emph{et~al.}) \emph{\bibinfo{booktitle}{Advances in Neural Information Processing Systems}}, Vol.~\bibinfo{volume}{30} (\bibinfo{publisher}{Curran Associates, Inc.}, \bibinfo{year}{2017}).

\bibitem{madani2023large}
\bibinfo{author}{Madani, A.} \emph{et~al.}
\newblock \bibinfo{title}{Large language models generate functional protein sequences across diverse families}.
\newblock \emph{\bibinfo{journal}{Nature Biotechnology}} \textbf{\bibinfo{volume}{41}}, \bibinfo{pages}{1099--1106} (\bibinfo{year}{2023}).

\bibitem{brandes2022proteinbert}
\bibinfo{author}{Brandes, N.}, \bibinfo{author}{Ofer, D.}, \bibinfo{author}{Peleg, Y.}, \bibinfo{author}{Rappoport, N.} \& \bibinfo{author}{Linial, M.}
\newblock \bibinfo{title}{Proteinbert: a universal deep-learning model of protein sequence and function}.
\newblock \emph{\bibinfo{journal}{Bioinformatics}} \textbf{\bibinfo{volume}{38}}, \bibinfo{pages}{2102--2110} (\bibinfo{year}{2022}).

\bibitem{rives2021biological}
\bibinfo{author}{Rives, A.} \emph{et~al.}
\newblock \bibinfo{title}{Biological structure and function emerge from scaling unsupervised learning to 250 million protein sequences}.
\newblock \emph{\bibinfo{journal}{Proceedings of the National Academy of Sciences}} \textbf{\bibinfo{volume}{118}}, \bibinfo{pages}{e2016239118} (\bibinfo{year}{2021}).

\bibitem{honda2020cross}
\bibinfo{author}{Honda, S.}, \bibinfo{author}{Koyama, K.} \& \bibinfo{author}{Kotaro, K.}
\newblock \emph{\bibinfo{title}{Cross attentive antibody-antigen interaction prediction with multi-task learning}} (\bibinfo{year}{2020}).

\bibitem{dens2023cross}
\bibinfo{author}{Dens, C.}, \bibinfo{author}{Laukens, K.}, \bibinfo{author}{Meysman, P.} \& \bibinfo{author}{Bittremieux, W.}
\newblock \bibinfo{title}{A cross-attention transformer encoder for paired sequence data}.
\newblock \emph{\bibinfo{journal}{bioRxiv}} \bibinfo{pages}{2023--12} (\bibinfo{year}{2023}).

\bibitem{miyato2017adversarial}
\bibinfo{author}{Miyato, T.}, \bibinfo{author}{Dai, A.~M.} \& \bibinfo{author}{Goodfellow, I.}
\newblock \emph{\bibinfo{title}{Adversarial training methods for semi-supervised text classification}} (\bibinfo{year}{2017}).

\bibitem{hui_2024_14282419}
\bibinfo{author}{Hui, L.}
\newblock \bibinfo{title}{A unified cross-attention model for predicting antigen binding specificity to both hla and tcr molecules} (\bibinfo{year}{2024}).
\newblock \urlprefix\url{https://doi.org/10.5281/zenodo.14282419}.

\end{thebibliography}

\end{document}